\documentstyle[preprint,aps]{revtex}

\tightenlines

\begin{document}

\newcommand{\tls}{C}
\newcommand{\cnb}{{\cal I}}
\newcommand{\ep}{\epsilon}
\newcommand{\Om}{\Omega}
\newcommand{\scri}{{\cal I}}
\newcommand{\lie}{{\pounds}}
\newcommand{\Fone}{F^1}
\newcommand{\Ftwo}{F^2}
\renewcommand{\FR}{F^{\rm R}}
\newcommand{\tFR}{\tilde{F}^{\rm R}}
\newcommand{\Fin}{F^{\rm in}}
\newcommand{\FA}{F^{\rm A}}
\newcommand{\tFA}{\tilde{F}^{\rm A}}
\newcommand{\Fbar}{\bar{F}}
\newcommand{\Ffree}{F^{\rm free}}
\newcommand{\tFfree}{\tilde{F}^{\rm free}}
\newcommand{\Frad}{F^{\rm rad}}
\newcommand{\gin}{\gamma^{\rm in}}
\newcommand{\gfree}{\gamma^{\rm free}}
\newcommand{\gA}{\gamma^{\rm A}}
\newcommand{\gR}{\gamma^{\rm R}}
\newcommand{\gtail}{\gamma^{\rm tail}}
\newcommand{\gbar}{\bar{\gamma}}
\newcommand{\Cfree}{C^{\rm free}}
\newcommand{\AR}{A^{\rm R}}
\renewcommand{\AA}{A^{\rm A}}
\newcommand{\Abar}{\bar{A}}
\newcommand{\GR}{G^{\rm R}}
\newcommand{\GA}{G^{\rm A}}
\newcommand{\Gbar}{\bar{G}}
\newcommand{\fg}{\hat{f}}
\newcommand{\reals}{{\cal R}}
\newcommand{\FS}{F^{\rm S}}
\newcommand{\tFS}{\tilde{F}^{\rm S}}
\newcommand{\TS}{T^{\rm S}}
\newcommand{\AS}{A^{\rm S}}
\newcommand{\jS}{j^{\rm S}}
\newcommand{\phit}{\tilde{\phi}}
\newcommand{\phiAt}{\tilde{\phi}^{\rm A}}
\newcommand{\phiRt}{\tilde{\phi}^{\rm R}}
\newcommand{\phiA}{\phi^{\rm A}}
\newcommand{\phiR}{\phi^{\rm R}}
\newcommand{\phifree}{\phi^{\rm free}}
\newcommand{\phitail}{\phi^{\rm tail}}
\newcommand{\phiin}{\phi^{\rm in}}
\newcommand{\nt}{\widetilde{\nabla}}
\newcommand{\gt}{\tilde{g}}
\newcommand{\ept}{\tilde{\epsilon}}
\newcommand{\surf}{{\cal S}}
\newcommand{\Vtail}{V^{\rm tail}}

\title{Energy conservation for point particles undergoing radiation
reaction}

\author{Theodore C. Quinn and Robert M. Wald}
\address{Enrico Fermi Institute and Department of Physics\\
University of Chicago\\
5640 S. Ellis Avenue\\
Chicago, Illinois 60637-1433}

\date{\today}

\maketitle

\begin{abstract}

For smooth solutions to Maxwell's equations sourced by a smooth
charge-current distribution $j_a$ in stationary, asymptotically flat
spacetimes, one can prove an energy conservation theorem which asserts
the vanishing of the sum of (i) the difference between the final and
initial electromagnetic self-energy of the charge distribution, (ii)
the net electromagnetic energy radiated to infinity (and/or into a
black hole/white hole), and (iii) the total work done by the
electromagnetic field on the charge distribution via the Lorentz
force. A similar conservation theorem can be proven for linearized
gravitational fields off of a stationary, asymptotically flat
background, with the second order Einstein tensor playing the role of
an effective stress-energy tensor of the linearized field. In this
paper, we prove the above theorems for smooth sources and then
investigate the extent to which they continue to hold for point
particle sources. The ``self-energy'' of point particles is ill
defined, but in the electromagnetic case, we can consider situations
where, initially and finally, the point charges are stationary and in
the same spatial position, so that the self-energy terms should
cancel. Under certain assumptions concerning the decay behavior of
source-free solutions to Maxwell's equations, we prove the vanishing
of the sum of the net energy radiated to infinity and the net work
done on the particle by the DeWitt-Brehme radiation reaction force. As
a by-product of this analysis, we provide a definition of the
``renormalized self-energy'' of a stationary point charge in a
stationary spacetime. We also obtain a similar conservation theorem
for angular momentum in an axisymmetric spacetime. In the
gravitational case, we argue that similar conservation results should
hold for freely falling point masses whose orbits begin and end at
infinity. This provides justification for the use of energy and
angular momentum conservation to compute the decay of orbits due to
radiation reaction. For completeness, the corresponding conservation
theorems for the case of a scalar field are given in an appendix.

\end{abstract}

\section{Introduction}
\label{intro}

The problem of calculating the motion of an isolated body coupled to
fields in curved spacetime is an old one which is currently receiving
renewed interest. This recent interest is largely driven by the need
for accurate calculations of processes which emit gravitational waves
(in anticipation of results from the new generation of gravitational
wave detectors) but similar issues arise for bodies coupled to
electromagnetic and scalar fields. In order to describe in a simple
manner those aspects of a body's motion which are independent of its
detailed internal structure, one often attempts to calculate the
motion of a ``point particle.'' A central problem in all such
investigations is to calculate the effects of the particle's own
fields, commonly referred to as ``self force'' or ``radiation
reaction'' effects. This problem is mathematically ill posed since the
fields diverge on the world line of the particle itself.

Nevertheless, there is a long history of attempts to calculate the
self-force on a particle directly from the local fields. All such
schemes involve some prescription for subtracting away the infinite
contributions to the force due to the singular nature of the field on
the particle's world line. In 1938, Dirac produced a force expression
for a point charge coupled to an electromagnetic field in Minkowski
spacetime by imposing local energy conservation on a tube surrounding
the particle's world line~\cite{dirac}. The infinite contributions to
the force were subtracted through a ``mass renormalization'' scheme.
In 1960, Dewitt and Brehme~\cite{dewittbrehme} generalized this
approach to an arbitrary curved background spacetime. (A trivial
calculational error in their paper was later corrected by
Hobbs~\cite{hobbs}.) More recently, Mino et al.~\cite{minoetal}
further adapted this approach to produce an expression for the self
force on a massive particle coupled to linearized gravity on a vacuum
background spacetime. Recently, we also have derived~\cite{quinnwald}
the formulas for the electromagnetic and gravitational self-forces by
using an axiomatic approach which, in effect, regularizes the forces
by comparing forces in different spacetimes.

Despite the fact that it is thereby known, in principle, how to
calculate the electromagnetic and gravitational self-force on a point
particle, serious difficulties arise in practice when one attempts to
evaluate this self-force on account of the difficulties in computing
the ``tail term'' contribution. Indeed, the evaluation of the ``tail
term'' is highly nontrivial even in the slow motion, weak field
limit~\cite{dewittdewitt,quinnwiseman}. Consequently, many researchers
have employed the following iterative strategy for calculating the
motion of point particles. First, one calculates the motion of the
particle in the absence of self-force effects. Then, the energy
(and/or angular momentum) radiated to infinity by the resulting fields
is calculated. Finally, the effects of this energy (and/or angular
momentum) loss are introduced as a perturbation to the particle's
motion. Note that the applicability of this approach is limited in
that the energy and angular momentum of the particle are not even
defined when the background spacetime fails, respectively, to be
stationary and axisymmetric. In addition, even when the energy and/or
angular momentum of the particle are well defined, the energy and angular momentum
radiated to infinity would be expected to equal the energy and angular
momentum loss by the particle only in a time averaged sense\footnote{The
example of a point charge in Minkowski spacetime which undergoes a
period of uniform acceleration explicitly demonstrates the failure of
temporally local conservation of this sort, since the radiation
reaction force (and, hence, the work done by it) vanishes at retarded
times during which a nonzero flux of radiation reaches
infinity. Nevertheless, the net work done by the radiation reaction
force equals the net energy radiated to infinity if the motion of the
particle is static at early and late times~\cite{jackson}. The
generalizations of such average conservation results to curved spacetime
and to the gravitational case are the main subjects of the present
paper.}  Hence, the approach is, in essence, limited to calculating
the secular decay of an otherwise stable orbit. Furthermore, in the absence of
spherical symmetry, the loss of energy and angular momentum in general is not
sufficient to determine even this secular decay (for example, it does
not determine the variation of the Carter constant for non-equatorial
orbits in the Kerr spacetime). Finally, this approach does not allow
one to calculate any so-called ``conservative forces'', i.e.,
contributions to the self-force which are not associated with energy
or angular momentum loss. Nevertheless, this approach is extremely
simple to apply and has been widely used to estimate the effects of
radiation reaction on the motion of point particles.

Although the derivations of the self-force given in~\cite{dirac},
\cite{dewittbrehme}, and~\cite{minoetal} were heuristically motivated
by local conservation of energy, it is far from obvious, a priori,
that they satisfy the property of ``global energy conservation'' as
assumed in the iterative procedure described above. Specifically, in
order to justify the iterative procedure, it is necessary that the
total energy and angular momentum radiated to infinity by the
electromagnetic or gravitational fields be equal, respectively, to the
net work and torque done by the self-force over the world line
of the particle. The main purpose of this paper is to investigate the
extent to which this is the case.

Our main results are the following: In the electromagnetic case, we
consider stationary, globally hyperbolic, asymptotically flat
spacetimes. We assume that source-free Maxwell fields satisfy a
certain decay property, and that the advanced and retarded solutions
with stationary sources have suitable fall-off at infinity. We then
prove that if a point charge is asymptotically stationary in the past
and future and in the same position, then the net electromagnetic
energy radiated to infinity (and/or into a black hole/white hole) is
equal to the integral over the particle's world line of the force
expression given by DeWitt and Brehme (as corrected by Hobbs)
contracted with the timelike Killing field. This provides some
justification for the use of the above global energy conservation
method to calculate the effects of radiation reaction. (Alternatively,
the fact that the DeWitt-Brehme force gives rise to global energy
conservation may be viewed as providing further justification for its
own validity.) As byproduct of this theorem, we show that it is
possible to consistently define the ``renormalized self-energy'' of a
stationary point charge. We argue that a similar energy conservation
result should also hold for a point charge which is in inertial motion
near infinity in the asymptotic past and future. We also show that
similar results hold for angular momentum conservation in axisymmetric
spacetimes. As an additional byproduct of our analysis, we show that
our conservation theorems also hold for the force expression used by
Gal'tsov~\cite{galtsov} and others, in which the Lorentz force
associated with the advanced-minus-retarded ``radiative'' field is
used. This result clarifies the relationship between the DeWitt-Brehme
and Gal'tsov formulas and demonstrates explicitly that global energy
conservation alone is insufficient to determine the local force.

Our analysis of the gravitational case closely parallels that of the
electromagnetic case, with the second-order Einstein
tensor~\cite{habisohn} of the metric perturbation replacing the
stress-energy tensor of the electromagnetic field. However, our
analysis is hampered by a number of technical obstacles---such as the
fact that the energy flux as calculated from the second order Einstein
tensor has been proven to agree with the Bondi flux only for
perturbations of compact spatial support---and its domain of
applicability is limited to freely falling point masses (since if the
point mass is not freely falling, an additional stress-energy source
for the metric perturbation must be present). Nevertheless, we argue
that energy and angular momentum conservation results analogous to the
those in the electromagnetic case should hold for a point mass which
is in geodesic motion near infinity in the asymptotic past and future.

The electromagnetic case is analyzed in Sec.~II. We begin by
proving energy conservation theorems for the case of smooth
charge-current distributions in a spacetime with no black or white
holes. The desired theorem for the case of point particle sources is
then proven. Generalizations to establish angular momentum
conservation and to allow for the presence of black and white holes
are then described.

Linearized gravitational perturbations are considered in
Sec.~III. Again, we begin by proving a conservation theorem for smooth
stress-energy sources and then analyze the point particle case.

Finally, for completeness, we present the analogous results for a
scalar field in an appendix.

Our notation and conventions throughout the paper follow~\cite{wald1}.

\section{Electromagnetic case}
\label{em}

In this section, we will state and prove our conservation theorems for
the case of a Maxwell field in a fixed background spacetime. For
simplicity, we shall first consider conservation of energy in a
stationary, asymptotically flat spacetime which contains no black or
white holes. Generalizations to other symmetries and to allow for the
presence of black holes and white holes will be discussed at the end
of this section. Thus, until these generalizations are considered, we
will restrict consideration to spacetimes satisfying the following
conditions:

\noindent {\bf Spacetime assumptions}: {\em Let $(M, g_{ab})$ be a
globally hyperbolic spacetime that is asymptotically flat at null and
spatial infinity in the sense of Ashtekar and
Hansen\cite{ashtekarhansen} and is stationary in the sense that it
possesses a Killing field $t^a$ that is asymptotically a time
translation at infinity. We further assume that no black holes or
white holes are present in $(M, g_{ab})$ (i.e., the domain of outer
communications is the entire spacetime) and that there exists a
smooth, spacelike Cauchy surface, $\Sigma$ for $(M, g_{ab})$ such that
in the unphysical spacetime $(\tilde{M}, \tilde{g}_{ab})$, $\Sigma
\cup i^0$ is compact. (This implies that $\Sigma$ is of the form of a disjoint union
$\Sigma = \Sigma_{\rm end} \cup \Sigma'$ where $\Sigma_{\rm end}$ has
the topology of $\reals^3$ minus a closed ball and $\Sigma'$ is
compact.)}

Without loss of generality, we may assume that $t^a n_a < 0$
everywhere on $\Sigma$, where $n^a$ is the future directed normal to
$\Sigma$ (see proposition 4.1 of~\cite{cw}).  We now deform $\Sigma$
to the future in a small neighborhood of $i^0$ so that the deformed
surface, $\tls^+$, satisfies $[\tls^+ \cap \Sigma] \supset \Sigma'$
and, in the unphysical spacetime, $\tls^+$ remains a smooth, spacelike
hypersurface, but now intersects $\scri^+$ in a cross-section ${\cal
S}^+$.\footnote{One way of doing this would be to choose a
cross-section ${\cal S}^+$ of $\scri^+$ such that $[J^-({\cal S}^+)
\cap \Sigma] \subset \Sigma_{\rm end}$ and consider the $C^0$,
partially null hypersurface $[\Sigma - I^-({\cal S}^+)] \cup
[J^-({\cal S}^+) \cap J^+(\Sigma)]$; then smooth this hypersurface
(see~\cite{seifert}) to a smooth spacelike hypersurface.} We similarly
deform $\Sigma$ to the past to construct a smooth, spacelike hypersurface
$\tls^-$ such that $[\tls^- \cap \Sigma] \supset \Sigma'$ and $\tls^-$
intersects $\scri^-$ in a cross-section ${\cal S}^-$. For all $t^+ >
0$, we define $\tls^+(t^+)$ to be the hypersurface obtained by ``time
translating'' $\tls^+$ along the orbits of $t^a$ by $t^+$. Similarly,
for all $t^- < 0$, we define $\tls^-(t^-)$ to be the hypersurface
obtained by ``time translating'' $\tls^-$ along the orbits of $t^a$ by
$t^-$. It then follows that for all $t^+ > 0$ and $t^- < 0$ the
surface $\tls^+(t^+) \cup [\scri^+ \cap J^-({\cal S}^+(t^+))] \cup
\tls^-(t^-) \cup [\scri^- \cap J^+({\cal S}^-(t^-))] \cup i^0$ bounds
a compact region $V(t^+, t^-)$, of the unphysical spacetime, as
illustrated in Figure~\ref{unphysical}.  Furthermore, it follows that
any $p \in M$ lies within $V(t^+, t^-)$ for sufficiently large $t^+,
t^-$.

In a spacetime $(M, g_{ab})$ satisfying the above properties, we wish
to consider solutions to Maxwell's equations with source $j_a$,
\begin{eqnarray}
\nabla^a F_{ab} &=& -4 \pi j_b \\
\nabla_{[a} F_{bc]} &=& 0.
\label{maxwell}
\end{eqnarray}
It follows immediately from Maxwell's equations that the stress-energy
tensor
\begin{equation}
T_{ab} = \frac{1}{4\pi} \left(F_{a}{}^c F_{bc}
                              - \frac{1}{4} g_{ab} F^{cd} F_{cd}\right)
\end{equation}
satisfies
\begin{equation}
\nabla^{b} T_{ab} = - F_{ab} j^b.
\label{Tcons}
\end{equation}
We define the stress-energy current three-form $J_{abc}$ by
\begin{equation}
J_{abc} = t^e T_{de} \epsilon_{abc}{}^d
\end{equation}
where $\epsilon_{abcd}$ is the (positively oriented) volume element
associated with the spacetime metric. We choose the orientation of
all spacelike and null hypersurfaces to be given by $v^a
\epsilon_{abcd}$, where $v^a$ is any future directed timelike vector
field. Then, the integral of $J_{abc}$ over a spacelike hypersurface
represents the total electromagnetic energy on that hypersurface,
whereas the integral of $J_{abc}$ over a null hypersurface represents
the flux of electromagnetic energy through that hypersurface. Equation
(\ref{Tcons}) implies that
\begin{eqnarray}
(dJ)_{abcd} &=& - \nabla^f (t^e T_{fe}) \epsilon_{abcd} \nonumber \\
&=& - t^e  \nabla^f T_{fe} \epsilon_{abcd} \nonumber \\
&=& t^e F_{ef} j^f \epsilon_{abcd}
\label{Jcons}
\end{eqnarray}

Initially, we shall consider smooth solutions to Maxwell's equations
with a smooth source $j_a$, but later in this section we will consider
solutions with point particle sources which are smooth away from the
world line of the particle. In all cases we shall consider only
solutions which satisfy the following three conditions:

\noindent {\bf Maxwell field assumptions}: {\em (1) $j_a$ vanishes in
a neighborhood of $\scri^+ \cup \scri^- \cup i^0$. (2) The unphysical
Maxwell field $\tilde{F}_{ab} = F_{ab}$ continuously extends to
$\scri^+$ and $\scri^-$. (3) The physical energy current three-form
$t^e T_{de} \epsilon_{abc}{}^d$ falls off sufficiently rapidly at
spatial infinity that (a) its integral over $\Sigma_{\rm end}$
is finite and (b) its integral over ${\cal B}_r \cap M$
vanishes as $r \rightarrow 0$, where ${\cal B}_r$ is a suitably chosen
coordinate sphere of radius $r$ around $i^0$ in the unphysical
spacetime.}

Condition (2) has the immediate consequence that the unphysical energy
current three-form $t^e \tilde{T}_{de} \tilde{\epsilon}_{abc}{}^d$
(where $\tilde{\epsilon}_{abc}{}^d$ is the metric-compatible volume
element associated with the unphysical metric, with index raised using
the unphysical metric) continuously extends to $\scri^+$ and
$\scri^-$. However, for a conformally invariant field such as
$F_{ab}$, the energy current three-form is conformally invariant with
conformal weight 0, so condition (2) implies that the physical energy
current three-form $t^e T_{de} \epsilon_{abc}{}^d$ also continuously
extends to $\scri^+$ and $\scri^-$.

Our fundamental result is simply a direct application of Stokes'
theorem to the integral of eq.~(\ref{Jcons})
over the region $V(t^+, t^-)$. Taking account of the absence of
contributions from spatial infinity as a consequence of assumption (3)
above (i.e., initially excluding from $V(t^+, t^-)$ a coordinate ball of radius $r$
around $i^0$ and then letting $r \rightarrow 0$), we obtain
\begin{eqnarray}
\int_{\tls^+(t^+)}
    \!\! t^a T_{ab} \epsilon_{cde}{}^b
  - \int_{\tls^-(t^-)}
    \!\! t^a T_{ab} \epsilon_{cde}{}^b && \nonumber \\
  \mbox{} + \int_{\scri^+ \cap J^-({\cal S}^+(t^+))} \!\! t^a T_{ab} \epsilon_{cde}{}^b
  - \int_{\scri^- \cap J^+({\cal S}^+(t^-))} \!\! t^a T_{ab} \epsilon_{cde}{}^b
  &=& \int_{V(t^+, t^-)} \!\! t^a F_{ab} j^b \epsilon_{cdef},
\label{stokes1}
\end{eqnarray}
Taking the limits as
$t^+ \rightarrow \infty$ and $t^- \rightarrow - \infty$, we obtain our
desired theorem for the case of smooth sources:

\noindent {\bf Theorem 2.1} Let $(M, g_{ab})$ satisfy the properties
stated above and let $F_{ab}$ be a smooth solution to Maxwell's
equations with smooth source $j_a$ which satisfies the three
properties stated above. Then,
\begin{eqnarray}
\lim_{t^+ \rightarrow +\infty} \int_{\tls^+(t^+)}
    \!\! t^a T_{ab} \epsilon_{cde}{}^b
  - \lim_{t^- \rightarrow -\infty} \int_{\tls^-(t^-)}
    \!\! t^a T_{ab} \epsilon_{cde}{}^b && \nonumber \\
  \mbox{} + \int_{\scri^+} \!\! t^a T_{ab} \epsilon_{cde}{}^b
  - \int_{\scri^-} \!\! t^a T_{ab} \epsilon_{cde}{}^b 
   &=& \int_{M} \!\! t^a F_{ab} j^b \epsilon_{cdef},
\label{smoothEM}
\end{eqnarray}
provided that each of the limits and integrals in the above formula exist.

\medskip

The above theorem has a straightforward physical interpretation. The
first pair of terms on the left side can be interpreted as being the
difference between the initial and final electromagnetic energy of the
charge-current distribution. The second pair of terms is just the net
electromagnetic energy radiated to infinity. Finally, since $F_{ab}
j^b$ is just the Lorentz force acting on the charge distribution, the
right side is the negative of the net work done by the electromagnetic field 
on the charge distribution, or, equivalently, it is the net work done by
the charge distribution on the electromagnetic field. This, in turn,
has the interpretation of representing the net amount of ``mechanical
energy'' which is converted to electromagnetic energy. Thus,
eq.~(\ref{smoothEM}) can be interpretated as stating that total energy
is conserved.

Unfortunately, this theorem as stated does not have a direct
counterpart for point particle sources. It is true that the net
electromagnetic energy radiated to infinity is perfectly well defined
in the limit of point particle sources. Furthermore, although the
integrand on the right side of eq.~(\ref{smoothEM}) becomes singular in the point
particle limit, one might nevertheless hope that, in a suitable
limiting process, the integral on the right side of
eq.~(\ref{smoothEM}) would converge to
\begin{equation}
\int \!\! t^a f_a \, d\tau
\end{equation}
where $f^a$ is the DeWitt and Brehme~\cite{dewittbrehme} expression
for the electromagnetic self-force on the particle (see
eq.~(\ref{DBforce}) below). However, the self-energy terms in
eq.~(\ref{smoothEM}) become hopelessly divergent in the point particle
limit, so, as it stands, eq.~(\ref{smoothEM}) will not make sense for
point particles.

Nevertheless, we can proceed by considering cases where the
self-energy terms in eq.~(\ref{smoothEM}) should cancel. In
particular, this should occur if at sufficiently early times the
charge-current source $j_a$ is stationary (i.e., $\lie_t j_a = 0$),
and at sufficiently late times, $j_a$ returns to the same stationary
state, i.e., if $j_a$ differs from a stationary distribution only in a
compact spacetime region. However, even in this case, in order to
ensure cancellation of the self-energy terms, it is necessary to
assume suitable decay properties of source free solutions, as well as
some properties of stationary solutions.

It order to formulate our decay assumptions, we need to introduce a
suitable norm, and it is most convenient to introduce this norm in
terms of the unphysical variables. On the hypersurface $\tls^+$, we
decompose the unphysical electromagnetic field $\tilde{F}_{ab} =
F_{ab}$ into its electric and magnetic parts with respect to the
unphysical unit normal $\tilde{n}^a = \Omega^{-1} n^a$ to $\tls^+$,
\begin{equation}
\tilde{E}_a = \tilde{F}_{ab} \tilde{n}^b
\label{E}
\end{equation}
\begin{equation}
\tilde{B}_a = - \frac{1}{2} \tilde{\epsilon}_{ab}{}^{cd} \tilde{F}_{cd} 
\tilde{n}^b
\label{B}
\end{equation}
On $\tls^+$, we define
\begin{equation}
\|F\| = \sup_{\tls^+}(\tilde{E}^a \tilde{E}_a 
+ \tilde{B}^a \tilde{B}_a)^{1/2}
\label{norm1}
\end{equation}
where the indices are raised and lowered with respect to the
unphysical metric. Note that for any $F_{ab}$ which is nonsingular in
the physical spacetime, we have $\|F\| < \infty$ provided only that
$\tilde{F}_{ab}$ continuously extends to $\scri^+$, as we have
assumed. Note also that since the components of $t^a$ are bounded in
an unphysical orthonormal frame associated with $\tilde{n}^a$, it
follows that everywhere on $\tls^+$ we have $|\tilde{T}_{ab} t^a
\tilde{n}^b| \leq C \|F\|^2$ for some constant $C$. Since $\tls^+$ has
finite volume with respect to the unphysical metric, it follows that
there exists a constant, $c$, such that
\begin{equation}
\int_{\tls^+} \!\!  t^a \tilde{T}_{ab} \tilde{\epsilon}_{cde}{}^b
\leq c \|F\|^2
\end{equation}
Since the energy current 3-form is conformally invariant, this is
equivalent to
\begin{equation}
\int_{\tls^+} \!\!  t^a T_{ab} \epsilon_{cde}{}^b
\leq c \|F\|^2
\label{enbnd}
\end{equation}
i.e., the norm we have introduced on $\tls^+$ bounds the energy on $\tls^+$.

In terms of the physical variables, we have
\begin{equation}
\|F\| = \sup_{\tls^+} \Omega^{-2} (E^a E_a + B^a B_a)^{1/2}
\label{norm2}
\end{equation}
where $E_a$ and $B_a$ are now defined with respect to the physical
unit normal $n^a$ and indices are now raised and lowered with respect
to the physical metric. We could similarly define a norm on
$\tls^+(t^+)$ by eqs.~(\ref{norm1}) or~(\ref{norm2}), but this would
not be useful on account of the possible ``time dependence'' of the
conformal factor $\Omega$. Thus, we instead define the function
$\Omega'$ on $\tls^+(t^+)$ by Lie transport of $\Omega$ along
$t^a$. (Equivalently, we could require that the conformal factor
$\Omega$ defining the unphysical spacetime be chosen so as to satisfy
$\lie_t \Omega = 0$ to the future of $\tls^+$.) For all $t^+ \geq
0$, we define
\begin{equation}
\|F\|(t^+) = \sup_{\tls^+(t^+)} (\Omega')^{-2} (E^a E_a + B^a B_a)^{1/2}
\label{normt+}
\end{equation}
We also define $\Omega'$ on all $\tls^-(t^-)$ by Lie transport of
$\Omega$ on $\tls^-$ and define $\|F\|(t^-)$ for all $t^- \leq 0$
similarly.

In the following, we shall restrict consideration to spacetimes which
satisfy the following ``decay hypothesis'':

\noindent {\bf Decay hypothesis}: {\em A spacetime $(M,g_{ab})$ satisfying the
conditions stated at the beginning of this section will be said to
satisfy the {\em decay hypothesis} for Maxwell fields if for every
smooth (in the physical spacetime) solution, $F_{ab}$, of the
source-free Maxwell equations which satisfies our Maxwell field
assumptions, we have}
\begin{equation}
\lim_{t^+ \rightarrow \infty} \|F\|(t^+) = \lim_{t^- \rightarrow
-\infty} \|F\|(t^-) = 0
\label{dh}
\end{equation}

It should be noted that although the definitions of $\|F\|(t^+)$ and
$\|F\|(t^-)$ depend upon the choice of conformal factor, $\Omega$, on
$\tls^+$ and $\tls^-$, it is clear that satisfaction of the decay
hypothesis does not depend upon this choice. Furthermore, it is not
difficult to verify that the satisfaction of the decay hypothesis does
not depend upon the choices of $\tls^+$ and $\tls^-$. Thus, the decay
hypothesis is, indeed, a condition on the spacetime, $(M,g_{ab})$.

If the Killing field $t^a$ is spacelike somewhere in $M$ (i.e., if an
ergoregion is present), then solutions of the source-free Maxwell
equations with negative total energy can be constructed, and since
only positive energy can be radiated to infinity, it is clear that the
decay hypothesis cannot hold (see~\cite{friedman}). However, we
conjecture that the decay hypothesis holds for all spacetimes
satisfying our spacetime assumptions in which $t^a$ is globally
timelike.

Note that if $(M,g_{ab})$ satisfies the decay hypothesis, then there
can exist at most one stationary solution to Maxwell's equations with
a stationary source $j_a$ which satisfies our Maxwell field
assumptions. (Proof: If two such solutions existed, their difference
would be a source-free solution which does not decay.) For a
stationary $j_a$, both the advanced and retarded solutions are
necessarily stationary. If $j_a$ is stationary and has compact spatial
support on a Cauchy surface, then condition (1) of the Maxwell field
assumptions automatically is satisfied. The final property we shall
need is that the advanced and retarded solutions associated with such
a $j_a$ satisfy conditions (2) and (3) as well:

\noindent {\bf Stationary solution property}: {\em $(M,g_{ab})$ will be said to
satisfy the {\em stationary solution property} if for any stationary
$j_a$ of compact spatial support, the advanced and retarded solutions
satisfy conditions (2) and (3) of the Maxwell field assumptions.}

It is readily verified that the stationary solution property holds for
Minkowski spacetime, and we believe that the stationary solution
property holds for all spacetimes satisfying our spacetime
assumptions. Indeed, near infinity, the behavior of stationary, source
free solutions is such that in order to prove that this property
holds, we would, in essence, need only show that the advanced and retarded
solutions for a stationary $j_a$ of compact spatial support do not
blow up at spatial infinity. While this seems undoubtedly true, one
would need some bounds on the long distance behavior of the advanced
and retarded Green's in a general spacetime satisfying our assumptions
in order to obtain a proof. 

Note that, in view of the above uniqueness property, if $(M,g_{ab})$
satisfies the decay hypothesis and the stationary solution property,
then for any stationary $j_a$ of compact spatial support, the advanced
and retarded solutions are equal. Note also that for any stationary
solution, the flux of energy through ${\cal I}^+$ between $i^0$ and
any cross-section ${\cal S}^+$ (or through ${\cal I}^-$ between $i^0$
and any cross-section ${\cal S}^-$) must either be zero or
infinite. The latter possibility is ruled out by our Maxwell field
assumptions. Thus, when the stationary solution property holds, the
retarded (= advanced) solution for any stationary $j_a$ of compact
spatial support must have vanishing energy flux through ${\cal I}^+$
and ${\cal I}^-$. From the form of the stress-energy tensor, it can be
seen that this is equivalent to the vanishing of the pullback to
${\cal I}^+$ (or ${\cal I}^-$) of $\tilde{F}_{ab} \tilde{n}^b$, where
$\tilde{n}^a$ denotes the normal to ${\cal I}^+$; equivalently, we
have on ${\cal I}^+$
\begin{equation}
\tilde{F}_{ab} \tilde{n}^b = \alpha \tilde{n}_a
\label{Fscri}
\end{equation}
for some function $\alpha$ on ${\cal I}^+$.

We have the following theorem:

\noindent {\bf Theorem 2.2}: Let $(M, g_{ab})$ be a spacetime satisfying the
conditions stated at the beginning of this section which, in addition,
satisfies the decay hypothesis and stationary solution property. Let
$F_{ab}$ be a smooth solution to Maxwell's equations satisfying our
Maxwell field assumptions such that, in addition, $j_a - \jS_a$
vanishes outside of a compact region of $M$, where $\jS_a$ is a
stationary charge-current distribution of compact spatial
support. Then, we have
\begin{equation}
\int_{\scri^+} \!\! t^a T_{ab} \epsilon_{cde}{}^b
  - \int_{\scri^-} \!\! t^a T_{ab} \epsilon_{cde}{}^b 
   = \int_{M} \!\! t^a F_{ab} j^b \epsilon_{cdef},
\label{smoothEM2}
\end{equation}

\noindent {\bf Proof}: Without loss of generality, we may assume that the
support of $\jS_a$ on $\Sigma$ is contained in $\Sigma'$, so that
$\jS_a$ vanishes in the region, $V(0,0)$, bounded by $C^+$ and
$C^-$.  Since eq.~(\ref{stokes1}) holds for $F_{ab}$, it suffices to
show that
\begin{equation}
\lim_{t^+ \rightarrow +\infty} \int_{\tls^+(t^+)} \!\!
t^a T_{ab} \epsilon_{cde}{}^b
\end{equation}
exists and equals
\begin{equation}
\lim_{t^- \rightarrow -\infty} \int_{\tls^-(t^-)} \!\!
t^a T_{ab} \epsilon_{cde}{}^b
\end{equation}
Let $\FS_{ab}$ be the retarded (= advanced) solution to Maxwell's
equations with source $\jS_a$, and let
\begin{equation}
{F'}_{ab} = F_{ab} - \FS_{ab}
\label{F'}
\end{equation}
Then, by straightforward estimates similar to those used to obtain
eq.~(\ref{enbnd}) above, we obtain for all $t^+ \geq 0$,
\begin{equation}
\left| \int_{\tls^+(t^+)} \!\!  t^a T_{ab} \epsilon_{cde}{}^b
- \int_{\tls^+(t^+)} \!\!  t^a \TS_{ab} \epsilon_{cde}{}^b \right|
\leq K \left[ \|F'\|(t^+) + \|\FS\|(t^+) \right] \|F'\|(t^+)
\end{equation}
for some constant $K$, where $\TS_{ab}$ denotes the stress-energy
tensor of $\FS_{ab}$.  However, by stationarity, we have
\begin{equation}
\int_{\tls^+(t^+)} \!\!  t^a \TS_{ab} \epsilon_{cde}{}^b
= \int_{\tls^+} \!\!  t^a \TS_{ab} \epsilon_{cde}{}^b
\end{equation}
Furthermore, since ${F'}_{ab}$ is source free for sufficiently large
$t^+$, the decay hypothesis applies to it. Consequently, we obtain
\begin{equation}
\lim_{t^+ \rightarrow +\infty} \int_{\tls^+(t^+)} \!\!  t^a T_{ab}
\epsilon_{cde}{}^b = \int_{\tls^+} \!\!  t^a \TS_{ab}
\epsilon_{cde}{}^b
\label{lim1}
\end{equation}
Similarly, we obtain
\begin{equation}
\lim_{t^- \rightarrow -\infty} \int_{\tls^-(t^-)} \!\!
t^a T_{ab} \epsilon_{cde}{}^b = \int_{\tls^-} \!\!
t^a \TS_{ab} \epsilon_{cde}{}^b
\label{lim2}
\end{equation}
Finally, we apply eq.~(\ref{stokes1}) to $\FS_{ab}$, choosing $t^+ =
t^- = 0$. Taking into account the fact that the energy flux integrals
through ${\cal I}^+$ and ${\cal I}^-$ vanish for $\FS_{ab}$ and that
$\jS_a$ vanishes in $V(0,0)$, we find that the right sides of
eqs.~(\ref{lim1}) and~(\ref{lim2}) are equal, as we desired to show.
\hfill \fbox{}

Applying this theorem to the case of a source-free Maxwell field
$F_{ab}$, we have 
\begin{equation}
\int_{\scri^+} \!\! t^a T_{ab} \epsilon_{cde}{}^b
  - \int_{\scri^-} \!\! t^a T_{ab} \epsilon_{cde}{}^b = 0.
\label{sf}
\end{equation}
That is, for a source-free solution, the energy radiated ``into'' the
spacetime through past null infinity is equal to the energy radiated
``out of'' the spacetime through future null infinity. This
observation, which will be important in our subsequent analysis,
demonstrates the intuitive content of our decay hypothesis: that no
energy remains ``trapped'' in the spacetime.

We are now prepared to generalize the above theorem to the case of a
point particle source with charge $e$ and world line $z(\tau)$. In
this case, the current $j_a$ is given by the distribution
\begin{equation}
j_a(x) = e \int\!\! \delta(x,z(\tau)) u_a \,d\tau .
\label{ppsource}
\end{equation}
Since Maxwell's equations are linear, they are well defined for
distributional sources. We shall consider only distributional
solutions $F_{ab}$ which are smooth away from the world line of the
particle and which satisfy the Maxwell field assumptions stated near
the beginning of this section. In that case, the integrals at
$\scri^+$ and $\scri^-$ appearing in eq.~(\ref{smoothEM2}) are well
defined.  However, since $F_{ab}$ is necessarily distributional in a
neighborhood of the world line of the particle the integral on the
right hand side of eq.~(\ref{smoothEM2}), which represents the work
done by the source on the field, contains a formal product of
distributions and is therefore ill defined. Nevertheless, we now wish
to show that, in the point particle case, we obtain a new version of
eq.~(\ref{smoothEM2}) in which the problematic product of
distributions on the right hand side is replaced by the integral
\begin{equation}
\int \!\! t^a f_a \, d\tau,
\end{equation}
where\footnote{The sign of the ``tail term'' in eq.~(19) of
reference~\cite{quinnwald} is incorrect due to an error in
transcribing DeWitt and Brehme's expression into our
notation. Additionally, a factor of 2 error was introduced in copying
the ``tail term'' to eq.~(23). As a result, the ``tail term'' in
eqs.~(25) and~(26) of reference~\cite{quinnwald} must be multiplied by
a factor of $-2$ in order to produce the correct formula (given in
eq.~(\ref{DBforce}) of the present paper).}
\begin{eqnarray}
f_a &=& e\Fin_{ab}u^b + \frac{2}{3} e^2 (\dot{a}_a - a^2 u_a) 
	+\frac{1}{3} e^2(R_{ab} u^b + u_a R_{bc}u^b u^c) \nonumber \\
&& + e^2u^b\int_{-\infty}^{\tau}
	2 \nabla_{[a} \GR_{b]c'} u^{c'}(\tau')\,d\tau',
\label{DBforce}
\end{eqnarray}
is the expression given by DeWitt and Brehme (as corrected by Hobbs)
for the total electromagnetic force on a point particle coupled to a
Maxwell field $F_{ab}$. In this force expression, $\GR_{ab}$ is the
retarded Green's function for the vector potential, satisfying
\begin{equation}
\nabla^b \nabla_b \GR_{aa'} - R_a{}^b \GR_{ba'} 
	= - 4 \pi \bar{g}_{aa'} \delta(x,z),
\end{equation}
$\FR_{ab}$ is the retarded solution with source (\ref{ppsource}), and
\begin{equation}
\Fin_{ab} \equiv F_{ab} - \FR_{ab} .
\end{equation}
In the integral over the particle's past world line, usually referred
to as the ``tail term'', there is an implicit limiting procedure: the
integral is performed from $-\infty$ to a point $\tilde{\tau} < \tau$
and then the limit $\tilde{\tau} \rightarrow \tau$ is taken. One
consequence of the Hadamard expansion is that this limit exists; i.e.,
no singular part of $G_{aa'}$ is encountered in this limit.

In order to better understand the behavior of eq.~(\ref{smoothEM2}) in
the point particle limit, we write $F_{ab}$ as the sum of the
half-advanced plus half retarded solution
\begin{equation}
\Fbar_{ab} \equiv (\FR_{ab} + \FA_{ab})/2
\end{equation}
and a source-free solution
\begin{equation}
\Ffree_{ab} \equiv F_{ab} - \Fbar_{ab} . 
\end{equation}
We shall consider only particle motions such that $\FR_{ab}$ and
$\FA_{ab}$ are smooth away from the world line of the particle and
such that $\FR_{ab}$ and $\FA_{ab}$ satisfy our Maxwell field
assumptions. If the stationary solution property holds, this will
hold whenever there exist $t^-, t^+$ such that the particle is
stationary for all $t < t^-$ and all $t > t^+$.

Since $T_{ab}$ is quadratic in $F_{ab}$, we have
\begin{equation}
T_{ab}[F,F] 
  = T_{ab}[\Fbar,\Fbar]
  + T_{ab}[\Ffree,\Ffree]
  + 2 T_{ab}[\Ffree,\Fbar],
\label{Fbreakup}
\end{equation}
where we have defined
\begin{equation}
T_{ab}[\Fone,\Ftwo] = \frac{1}{4\pi} \left(\Fone_{(a}{}^c \Ftwo_{b)c}
                      - \frac{1}{4} g_{ab} (\Fone)^{cd} \Ftwo_{cd}\right).
\end{equation}
Now, $\Ffree_{ab}$ is a source-free solution which is smooth away from
the world line of the particle and consequently (by the propagation
of singularities theorem~\cite {hormander}) is smooth everywhere.
Since $\Ffree_{ab}$ also satisfies our Maxwell field assumptions,
eq.~(\ref{sf}) applies to it, and $T_{ab}[\Ffree,\Ffree]$ will make zero
contribution to the net energy radiated to infinity (the left hand
side of eq.~(\ref{smoothEM2}) above). Furthermore, since
$T_{ab}[\Ffree,\Fbar]$ is the product of a distribution and a smooth
tensor field and thus is well defined as a distribution, this term
will give rise to a well-defined volume integral when we apply Stokes'
theorem; we will evaluate this term in Proposition 2.1 below. Thus,
the only term that is mathematically problematic is
$T_{ab}[\Fbar,\Fbar]$, which contains a product of
distributions. However, we will now show that, if the point particle
source is initially and finally stationary (i.e., if there exist $t^-,
t^+$ such that the particle is stationary for all $t < t^-$ and all $t
> t^+$), this term makes vanishing contribution to the net energy
radiated to infinity.

To show this, we write $T_{ab}[\Fbar,\Fbar]$ as
\begin{equation}
T_{ab}[\Fbar,\Fbar] = T_{ab}[\Frad,\Frad] + T_{ab}[\FR,\FA],
\end{equation}
where
\begin{equation}
\Frad_{ab} = (\FR_{ab} - \FA_{ab})/2 .
\end{equation}
Like $\Ffree_{ab}$, the radiative solution $\Frad_{ab}$ is a source-free
solution which is smooth away from the world line of the particle and,
consequently, is smooth everywhere. Since $\Frad_{ab}$ also satisfies
our Maxwell field assumptions, $T_{ab}[\Frad,\Frad]$ makes no
contribution to the net energy radiated to infinity. Therefore, we
have
\begin{equation}
\int_{\scri^+} \!\! t^a T_{ab}[\Fbar,\Fbar] \epsilon_{cde}{}^b
  - \int_{\scri^-} \!\! t^a T_{ab}[\Fbar,\Fbar] \epsilon_{cde}{}^b
= \int_{\scri^+} \!\! t^a T_{ab}[\FR,\FA] \epsilon_{cde}{}^b
  - \int_{\scri^-} \!\! t^a T_{ab}[\FR,\FA] \epsilon_{cde}{}^b
\label{CTvanish}
\end{equation}
However, since the particle is initially stationary, it follows that
$\FR_{ab}$ satisfies eq.~(\ref{Fscri}) on ${\cal I}^-$. Taking into
account the fact that on ${\cal I}^-$, $t^a$ is proportional to the
normal, $\tilde{n}^a$, to ${\cal I}^-$ and $\FA_{ab}$ is antisymmetric
in its indices, we see that the flux integral over ${\cal I}^-$ of the
``advanced-retarded cross-term'' vanishes. Similarly, using the fact
that the particle is finally stationary, we see that the integral of
this cross term over ${\cal I}^+$ also vanishes, which yields the
desired result.

We are now ready to state and prove the following intermediate
result\footnote{It should be noted that a direct analog of this
proposition also holds in the case of smooth sources and, indeed, the
present proposition could be obtained by taking the point particle
limit of the analogous result for smooth sources. We chose not to
present this result in the smooth case because Theorem 2.2 (i.e., the
smooth source analog of Theorem 2.3 below) could be proven more directly
by other means.}:

\noindent {\bf Proposition 2.1}: Let $(M,g_{ab})$ be a spacetime satisfying the
conditions stated at the beginning of this section together with the
decay hypothesis and stationary solution property. Let $z(\tau)$ be a
timelike curve which differs from an orbit, $z_0 (\tau)$, of the
stationary Killing field $t^a$ only over a finite interval. Let
$F_{ab}$ be a solution to Maxwell's equations with source
(\ref{ppsource}) which satisfies our Maxwell field assumptions. Then,
we have
\begin{equation}
\int_{\scri^+} \!\! t^a T_{ab} \epsilon_{cde}{}^b
  - \int_{\scri^-} \!\! t^a T_{ab} \epsilon_{cde}{}^b
  =  \int t^a \fg_a \, d\tau,
\label{propeq}
\end{equation}
where
\begin{equation}
\fg_a \equiv e \Ffree_{ab} u^b
\label{fg1}
\end{equation}

\noindent {\bf Proof}: We apply Stokes' theorem to the differential of energy
current three-form associated with the cross-term $2
T_{ab}[\Ffree,\Fbar]$. In analogy with the derivation of
eq.~(\ref{smoothEM}), we obtain
\begin{eqnarray}
\lim_{t^+ \rightarrow +\infty} 2 \int_{\tls^+(t^+)}
    \!\! t^a T_{ab}[\Ffree,\Fbar] \epsilon_{cde}{}^b
  - \lim_{t^- \rightarrow -\infty} 2 \int_{\tls^-(t^-)}
    \!\! t^a T_{ab}[\Ffree,\Fbar] \epsilon_{cde}{}^b && \nonumber \\
  \mbox{} + 2 \int_{\scri^+} \!\! t^a T_{ab}[\Ffree,\Fbar] \epsilon_{cde}{}^b
      - 2 \int_{\scri^-} \!\! t^a T_{ab}[\Ffree,\Fbar] \epsilon_{cde}{}^b
  &=& - 2 \int_M t^a \nabla^b T_{ab}[\Ffree,\Fbar] \epsilon_{cdef}
\nonumber \\
  &=& \int_M t^a \Ffree_{ab} j^b \epsilon_{cdef}
\nonumber \\
  &=& e \int t^a \Ffree_{ab} u^b \, d\tau
\nonumber \\
  &=& \int t^a \fg_a \, d\tau,
\end{eqnarray}
However, by the above results, we have
\begin{equation}
2 \int_{\scri^+} \!\! t^a T_{ab}[\Ffree,\Fbar] \epsilon_{cde}{}^b
      - 2 \int_{\scri^-} \!\! t^a T_{ab}[\Ffree,\Fbar] \epsilon_{cde}{}^b
= \int_{\scri^+} \!\! t^a T_{ab} \epsilon_{cde}{}^b
  - \int_{\scri^-} \!\! t^a T_{ab} \epsilon_{cde}{}^b
\end{equation}
Thus, it suffices to show that 
\begin{equation}
\lim_{t^+ \rightarrow +\infty} \int_{\tls^+(t^+)}
    \!\! t^a T_{ab}[\Ffree,\Fbar] \epsilon_{cde}{}^b =
\lim_{t^- \rightarrow -\infty} \int_{\tls^-(t^-)}
    \!\! t^a T_{ab}[\Ffree,\Fbar] \epsilon_{cde}{}^b = 0
\label{xlims}
\end{equation}
To show this, in analogy with eq.~(\ref{F'}) we define
\begin{equation}
{F''}_{ab} = \Fbar_{ab} - \FS_{ab}
\label{F''}
\end{equation}
where $\FS_{ab}$ is the stationary solution sourced by a point
charge following the orbit, $z_0 (\tau)$, of $t^a$. We use
eq.~(\ref{F''}) to substitute for $\Fbar$ in eq.~(\ref{xlims}). Since
$\Ffree$ is a source free solution satisfying our Maxwell field
assumptions and, in the asymptotic past and future, so is $F''$, it
follows immediately from the decay hypothesis that
\begin{equation}
\lim_{t^+ \rightarrow +\infty} \int_{\tls^+(t^+)}
    \!\! t^a T_{ab}[\Ffree,F''] \epsilon_{cde}{}^b =
\lim_{t^- \rightarrow -\infty} \int_{\tls^-(t^-)}
    \!\! t^a T_{ab}[\Ffree,F''] \epsilon_{cde}{}^b = 0
\end{equation}
On the other hand, although $\FS$ is singular on the world line of the
source, the Hadamard expansion shows that on $\tls^+(t^+)$ it diverges
as $1/\sigma$ as one approaches the point charge, where $\sigma$
denotes squared geodesic distance. Consequently, the unphysical
orthonormal frame components of $\tFS_{ab}$ are $L^1$
functions on $\tls^+(t^+)$ with respect to the unphysical volume
element $\tilde{\epsilon}_{abcd} \tilde{n}^a$. Therefore, we obtain
\begin{eqnarray}
\left| \int_{\tls^+(t^+)} \!\! t^a T_{ab}[\Ffree,\FS]
                               \epsilon_{cde}{}^b \right|
&=& \left| \int_{\tls^+(t^+)} \!\! t^a \tilde{T}_{ab}[\tFfree,\tFS]
\tilde{\epsilon}_{cde}{}^b \right| \nonumber \\
& \leq & k \|\FS\|_{L^1} \|\Ffree\|
\end{eqnarray}
for some constant $k$, which, together with the decay hypothesis
applied to $\Ffree$, implies
\begin{equation}
\lim_{t^+ \rightarrow +\infty} \int_{\tls^+(t^+)}
    \!\! t^a T_{ab}[\Ffree,\FS] \epsilon_{cde}{}^b = 0
\end{equation}
and similarly for the limit as $t^- \rightarrow - \infty$. \hfill \fbox{}

\medskip

The above proposition is a somewhat curious result, because although
it provides an expression of the general form we are seeking, the
force (\ref{fg1}) is {\it not} the force expression given by DeWitt and
Brehme (Eq.~(\ref{DBforce}) above). Indeed, we have
\begin{eqnarray}
\fg_a &=& e\Fin_{ab}u^b + \frac{2}{3} e^2 (\dot{a}_a - a^2 u_a) 
	+\frac{1}{3} e^2(R_{ab} u^b + u_a R_{bc}u^b u^c) \nonumber \\
&& + \frac{1}{2} e^2u^b\int_{-\infty}^{\tau}
	2\nabla_{[a} \GR_{b]c'} u^{c'}(\tau')\,d\tau'
   - \frac{1}{2} e^2u^b\int_{\tau}^{+\infty}
	2\nabla_{[a} \GA_{b]c'} u^{c'}(\tau')\,d\tau',
\label{Gforce}
\end{eqnarray}
which differs from eq.~(\ref{DBforce}) by the form of the tail
term. Indeed, for the retarded solution (so that $\Fin$ vanishes),
eq.~(\ref{Gforce}) contains acausal contributions from the future
history of the particle, whereas eq.~(\ref{DBforce}) does not. Thus,
although $\fg_a$ is the correct force expression in Minkowski
spacetime~\cite{dirac} (where the tail term vanishes) and it has been
used in curved spacetime, most notably by Gal'tsov~\cite{galtsov},
apparently because of the calculational simplicity it affords, it is
not equivalent to $f_a$. Indeed, we always have $\fg_a=0$ whenever
$\Ffree_{ab}=0$, so that, in particular, $\fg_a=0$ for a static charge
in Schwarzschild if there is no incoming radiation. On the other hand,
it is not difficult to verify that the force prescription given by
Smith and Will~\cite{smithwill} for a static charge in Schwarzschild
satisfies the axioms of~\cite{quinnwald} and thus agrees with the
DeWitt-Brehme force. The force calculated by Smith and Will is
nonvanishing.

We now calculate the difference between the work done by $f_a$ and
$\fg_a$. The difference between $f_a$ and $\fg_a$ is given by
\begin{eqnarray}
f_a - \fg_a &=& \frac{1}{2} e^2u^b\int_{-\infty}^{\tau}
	        2\nabla_{[a} \GR_{b]c'} u^{c'}(\tau')\,d\tau'
              + \frac{1}{2} e^2u^b\int_{\tau}^{+\infty}
	        2\nabla_{[a} \GA_{b]c'} u^{c'}(\tau')\,d\tau'
\nonumber \\
  &=& e^2u^b\int_{-\infty}^{+\infty}
	        2 \nabla_{[a} \Gbar_{b]c'} u^{c'}(\tau')\,d\tau',
\end{eqnarray}
where
\begin{equation}
\Gbar_{ab'} = (\GA_{ab'} + \GR_{ab'})/2
\end{equation}
and where it is understood that the singular point $\tau'=\tau$ is
omitted from the region of integration, i.e., one excludes an interval
of size $\epsilon$ centered at $\tau$ and then takes the limit as
$\epsilon \rightarrow 0$. Therefore, the difference between the total
work done by $f^a$ and the total work done by $\fg_a$ during the time
interval $[\tau^-, \tau^+]$ is\footnote{Note that we proceed by
considering the work done in the finite time interval $[\tau^-,
\tau^+]$ and only at the end of the calculation do we take the limit
$\tau^\pm \rightarrow \pm\infty$. We do so in order that all of the
integrals arising in our calculations will converge suitably well to
justify the interchanges of orders of integration.}
\begin{eqnarray}
\int_{\tau^-}^{\tau^+} t^a (f_a-\fg_a) \, d\tau
  &=& e^2 \int_{\tau^-}^{\tau^+} \!\! t^a
      \left[\int_{-\infty}^{\infty} 2\nabla_{[a}\Gbar_{b]b'} u^{b'}
            \, d\tau' \right]
      u^b \,d\tau
\nonumber \\
  &=& e^2 \int_{\tau^-}^{\tau^+} \!\!
      \int_{-\infty}^{\infty} 2 t^a \nabla_{[a}\Gbar_{b]b'} u^{b} u^{b'}
      \,d\tau' \,d\tau
\nonumber \\
  &=& - e^2 \int_{\tau^-}^{\tau^+} \!\! \int_{-\infty}^{\infty}
            u^b \nabla_b (t^a \Gbar_{ab'}) u^{b'} \,d\tau'\,d\tau
      + e^2 \int_{\tau^-}^{\tau^+} \!\! \int_{-\infty}^{\infty}
            \lie_{t}\Gbar_{bb'} u^{b} u^{b'} \, d\tau' \, d\tau.
\label{workdiff}
\end{eqnarray}
We now analyze the contributions of the two terms appearing on the
right side of eq.~(\ref{workdiff}).

Consider the first term. On account of the finite range of the $\tau$
integration, we may interchange the orders of integration to obtain
\begin{eqnarray}
e^2 \int_{\tau^-}^{\tau^+} \!\! \int_{-\infty}^{\infty}
  u^b \nabla_b (t^a \Gbar_{ab'})
  u^{b'} \,d\tau'\,d\tau
	&=& e^2 \int_{-\infty}^{\infty} 
                [t^a \Gbar_{ab'}(\tau^+, \tau') - 
	         t^a \Gbar_{ab'}(\tau^-, \tau')] u^{b'} d \tau' \nonumber \\
	&=& e \left[ t^a \Abar^{\rm tail}_a(z(\tau^+))
                    -t^a \Abar^{\rm tail}_a(z(\tau^-))\right],
\label{Gdiff}
\end{eqnarray}
where $\Abar^{\rm tail}_a$ is the ``tail part'' of the time-symmetric
vector potential for our point particle source. (Recall that the point
$\tau' = \tau^\pm$ is omitted from the integration in
eq.~(\ref{Gdiff}) so that $\Abar^{\rm tail}_a$ is finite.) Now, since
the particle is being held stationary for early and late times, the
solution must become stationary at early and late times,
so\footnote{Strictly speaking, the decay
hypothesis directly tells us only about the decay of the Maxwell field
tensor, $F_{ab}$. In order to obtain eq.~(\ref{Atail}), we must extend
the decay hypothesis to include the assumption that in the Lorentz
gauge, the vector potential of any smooth, source free solution goes
to zero at late times (say, pointwise along every Killing
orbit).\label{extdec}}
\begin{equation}
\lim_{\tau^\pm \rightarrow \pm\infty} t^a \Abar^{\rm
	tail}_a(z(\tau^\pm)) = - \Vtail({\bf x}^\pm, {\bf x}^\pm),
\label{Atail}
\end{equation}
where $\Vtail({\bf x}_1,{\bf x}_2)$ is the tail part of $- t^a
A_a$ as measured at spatial position ${\bf x}_1$ for a point particle
source held stationary at spatial position ${\bf x}_2$ for all time
and where ${\bf x}^{\pm} = \lim_{\tau^\pm \rightarrow \pm\infty} {\bf
x}(\tau^\pm)$. Therefore, we have for the first term
\begin{equation}
\lim_{\tau^\pm \rightarrow \pm\infty} 
  e^2 \int_{\tau^-}^{\tau^+} \!\! \int_{-\infty}^{\infty}
  u^b \nabla_b (t^a \Gbar_{ab'})
  u^{b'} \,d\tau'\,d\tau
  	= - e \Vtail({\bf x}^+, {\bf x}^+)
          + e \Vtail({\bf x}^-, {\bf x}^-).
\label{Vdiff}
\end{equation}
Since we further assume that the particle begins and ends in the {\em
same\/} spatial position, we have ${\bf x}^+ = {\bf x}^-$
and the right side vanishes.

On the other hand, the second term in eq.~(\ref{workdiff}) can be
analyzed as follows. The Green's function $\Gbar$ satisfies the
property that it is symmetric in its arguments,\footnote{This fact
follows from the antisymmetry of $\GA - \GR$ together with the support
properties of $\GA$ and $\GR$. The antisymmetry of $\GA - \GR$ follows
directly from the generalization to curved spacetime of the Maxwell
version of lemma 3.2.1 of~\cite{wald2} (see also
\cite{dewittbrehme}).}
\begin{equation}
\Gbar_{aa'}(x,x')=\Gbar_{a'a}(x',x).
\label{sym1}
\end{equation}
Furthermore, the invariance of $\Gbar$ under the time translation
isometries implies that
\begin{equation}
\lie_t \Gbar_{aa'} + \lie_{t'} \Gbar_{aa'} = 0 .
\label{sym2}
\end{equation}
Consequently, if we integrate over the {\it same} limits in $\tau$ and
$\tau'$, we have
\begin{eqnarray}
\int \int
 \lie_{t}\Gbar_{bb'}(\tau, \tau') u^{b} u^{b'} \,
  d\tau' \, d\tau
&=& \int \int
 \lie_{t'}\Gbar_{b'b}(\tau',\tau) u^{b} u^{b'} \,
  d\tau' \, d\tau \nonumber \\
&=& \int \int
 \lie_{t'}\Gbar_{bb'}(\tau,\tau') u^{b} u^{b'} \,
  d\tau' \, d\tau \nonumber \\
&=& - \int \int
 \lie_{t}\Gbar_{bb'}(\tau,\tau') u^{b} u^{b'} \,
  d\tau' \, d\tau
\label{symint}
\end{eqnarray}
Here the first equality was obtained by interchanging the dummy
variables $\tau$ and $\tau'$, the second equality was obtained using
eq.~(\ref{sym1}), and the last equality was obtained using
eq.~(\ref{sym2}). This shows that any integral of the form
(\ref{symint}) vanishes, provided only that the domain of integration
is symmetric in $\tau$ and $\tau'$ and that the integral is
suitably convergent in this domain to justify the interchange of
orders of integration.

Applying eq.~(\ref{symint}) to the domain $[\tau_-, \tau_+] \times
[\tau_-, \tau_+]$, we find that the second term in
eq.~(\ref{workdiff}) becomes
\begin{equation}
e^2 \int_{\tau^-}^{\tau^+} \!\!  \int_{-\infty}^{\infty}
  \lie_{t}\Gbar_{bb'} u^{b} u^{b'} \, d\tau' \, d\tau
  = e^2 \int_{\tau^-}^{\tau^+} \!\!  \int_{\tau^+}^{\infty}
        \lie_{t}\Gbar_{bb'} u^{b} u^{b'} \, d\tau' \, d\tau
   +e^2 \int_{\tau^-}^{\tau^+} \!\!  \int_{-\infty}^{\tau_-}
        \lie_{t}\Gbar_{bb'} u^{b} u^{b'} \, d\tau' \, d\tau
\end{equation}
Focusing attention on the first term, we find
\begin{eqnarray}
\lim_{\tau^+ \rightarrow +\infty}
  e^2 \int_{\tau^-}^{\tau^+} \!\!  \int_{\tau^+}^{\infty}
  \lie_{t}\Gbar_{bb'} u^{b} u^{b'} \, d\tau' \, d\tau
  &=& - \lim_{\tau^+ \rightarrow +\infty}
        e^2 \int_{\tau^-}^{\tau^+} \!\!  \int_{\tau^+}^{\infty}
        \lie_{t'}\Gbar_{bb'} u^{b} u^{b'} \, d\tau' \, d\tau \nonumber \\
  &=& - \lim_{\tau^+ \rightarrow +\infty}
        e^2 \int_{\tau^-}^{\tau^+} \!\!
        \left[\int_{\tau^+}^{\infty}
              \lie_{t'}(u^{b'} \Gbar_{bb'}) \, d\tau' \right]
        u^b \, d\tau \nonumber \\
  &=& \lim_{\tau^+ \rightarrow +\infty}
      e^2 \int_{\tau^-}^{\tau^+} \!\!
      \chi u^{b'} \Gbar_{bb'}(z(\tau),z(\tau^+))u^b \, d\tau \nonumber \\
  &=& \lim_{\tau^+ \rightarrow +\infty}
      \frac{e}{2} \chi u^{b'} \AA_{b'}(\tau^+) \nonumber \\
  &=& - \frac{e}{2} \Vtail({\bf x}^+,{\bf x}^+),
\end{eqnarray}
where we have used the fact that $t^a = \chi u^a$ for sufficiently
late times. Therefore we have
\begin{equation}
\lim_{\tau^+ \rightarrow +\infty}
  e^2 \int_{\tau^-}^{\tau^+} \!\!
  \int_{-\infty}^{\infty} \lie_{t}\Gbar_{bb'} u^{b} u^{b'} \, d\tau'
  \, d\tau = - \frac{e}{2} \left[ \Vtail({\bf x}^+, {\bf x}^+)
          - \Vtail({\bf x}^-, {\bf x}^-) \right]
\label{Lieterm}
\end{equation}
Again, this term vanishes when ${\bf x}^+ = {\bf x}^-$.

The above result together with the previous proposition yield the main
theorem of this section:

\noindent {\bf Theorem 2.3} Let $(M,g_{ab})$ be a spacetime satisfying
the conditions stated at the beginning of this section together with
the decay hypothesis (and its extension indicated in footnote
\ref{extdec}) and stationary solution property. Let $z(\tau)$ be a
timelike curve which differs from an orbit, $z_0 (\tau)$, of the
stationary Killing field $t^a$ only over a finite interval. Let
$F_{ab}$ be a solution to Maxwell's equations with source
(\ref{ppsource}) which satisfies our Maxwell field assumptions. Then,
we have
\begin{equation}
\int_{\scri^+} \!\! t^a T_{ab} \epsilon_{cde}{}^b
  - \int_{\scri^-} \!\! t^a T_{ab} \epsilon_{cde}{}^b
  =   \int t^a f_a \, d\tau.
\label{pointEM}
\end{equation}
where $f^a$ is the DeWitt-Brehme force, eq.~(\ref{DBforce}).

\medskip

Thus, both the correct force prescription $f_a$ and the incorrect
force prescription $\fg_a$ exhibit the property of global energy
conservation This underscores the fact that global energy conservation
is insufficient to determine a local expression for the force on a
point particle.

In Proposition 2.1 and Theorem 2.3, it was required that the
particle begin and end in the {\it same} stationary state. It is
interesting to consider the case where the particle is stationary for
$t < t^-$ and again becomes stationary for $t > t^+$, but its final
position differs from its initial position. In this case the proof of
Proposition 2.1 holds without essential change, and eq.~(\ref{propeq})
still applies. From eqs.~(\ref{workdiff}), (\ref{Vdiff}), and (\ref{Lieterm}),
we obtain
\begin{equation}
\int_{\scri^+} \!\! t^a T_{ab} \epsilon_{cde}{}^b
  - \int_{\scri^-} \!\! t^a T_{ab} \epsilon_{cde}{}^b
  + \frac{e}{2} \Vtail ({\bf x}^+,{\bf x}^+) - \frac{e}{2} \Vtail({\bf x}^-,{\bf x}^-)
	=  \int t^a f_a \, d\tau.
\label{pointEM'}
\end{equation}
Comparison with eq.~(\ref{smoothEM}) strongly suggests that we
identify $\frac{e}{2} \Vtail ({\bf x},{\bf x})$ as the {\it
renormalized electromagnetic self-energy}, $E_{\rm self}$, of a
stationary point charge at position ${\bf x}$
\begin{equation}
E_{\rm self} ({\bf x}) \equiv \frac{e}{2} \Vtail ({\bf x},{\bf x})
\label{Eself}
\end{equation}
From the analysis of Smith and Will~\cite{smithwill}, it
can be seen that for a static point charge in Schwarzschild spacetime,
we have
\begin{equation}
E_{\rm self} = e^2 M/2r^2 
\label{selfen}
\end{equation}
where $M$ is the mass of the Schwarzschild spacetime and $r$ is the
Schwarzschild radial coordinate.  For sufficiently slow motion
(relative to the static Killing field), the static force associated
with eq.~(\ref{selfen}) should dominate over any ``damping force''
associated with radiated energy\footnote{The static force associated
with eq.~(\ref{selfen}) is of order $e^2 M/r^3$, whereas in the slow
motion, weak field limit, the damping force on a circular geodesic
orbit is of order $e^2 \omega^3 r = e^2 M^{3/2}/r^{9/2}$ (see
\cite{dewittdewitt}), which is smaller by a factor of $(M/r)^{1/2}$.}.
Therefore, in the slow motion limit, we would expect to obtain the
dominant self-force correction to the motion of a freely falling test
charge in Schwarzschild spacetime by replacing the usual expression
for energy,
\begin{equation}
E \equiv - m g_{ab} t^a u^b = m \left(1 - \frac{2M}{r}\right) \dot{t},
\label{energydef}
\end{equation}
by
\begin{equation}
E' = E + E_{\rm self}.
\label{modenergydef}
\end{equation}
since $E'$ (rather than $E$) should be a constant of the motion.
Solving for $\dot{t}$ and plugging into the expression
$g_{ab} u^a u^b = -1$ with $L= m r^2 \dot{\phi}$, we obtain
\begin{equation}
\frac{1}{2} m \dot{r}^2 + \left(1-\frac{2M}{r}\right)
                        \left(\frac{L^2}{2mr^2} + \frac{m}{2}\right)
                      + \frac{E_{\rm self} E'}{m}
                      - \frac{E_{\rm self}^2}{2m}
                      = \frac{E'^2}{2m},
\end{equation}
which is the equation of motion for a particle of mass $m$ and energy
$E'^2/2m$ in a one-dimensional potential, $m U_{\rm eff}$. To lowest
nontrivial order in $e$, $U_{\rm eff}$ is given by
\begin{equation}
U_{\rm eff} = \frac{1}{2} - \frac{M}{r} + \frac{L^2}{2m^2r^2} -
  \frac{ML^2}{m^2 r^3} + \frac{e^2ME}{2m^2r^2}.
\label{Ueff}
\end{equation}
The last term in eq.~(\ref{Ueff}) represents the ``self-force''
correction to the motion of a charged particle in Schwarzschild
spacetime. Interestingly, for the case of a nearly extreme
Reissner-Nordstrom black hole, a self-energy correction of the
form~(\ref{modenergydef}) with $E_{\rm self}$ given by
eq.~(\ref{selfen}) is of just the right nature to eliminate the
possible counterexamples to cosmic censorship recently proposed by
Hubeny~\cite{hubeny}.

How are our arguments and results modified if the particle motion is
not stationary in the past and/or future? In our arguments above, we
used stationarity in the past and future only to obtain the following
four results: (1) To conclude (in conjunction with our stationary
solution property) that the advanced and retarded solutions satisfy
our Maxwell field assumptions. (2) To conclude that the
``advanced-retarded cross-term'' makes no contribution to the energy
flux through ${\cal I}^+$ and ${\cal I}^-$ (see the right side of
eq.~(\ref{CTvanish}) above). (3) To prove eq.~(\ref{xlims}). (4) To
evaluate the right side of eq.~(\ref{workdiff}). Thus,
eq.~(\ref{pointEM}) will continue to hold in all other circumstances
where the above results are valid and where the right side of
eq.~(\ref{workdiff}) vanishes. In particular, if we consider a
particle orbit which comes in from infinity in the asymptotic past in
(nearly) geodesic motion and emerges to infinity in the future also in
(nearly) geodesic motion, then, by calculations similar to those done
in the stationary case, the right side of eq.~(\ref{workdiff}) can be
shown to be given by the limit of ``tail terms'' in the asymptotic
past and future, which should vanish. However, it is somewhat more
delicate as to whether the other results hold. For example, results
(1)-(3) should hold for a point charge in Minkowski spacetime which
undergoes exactly inertial motion in the asymptotic past and future.
However, for the case of two point charges in Minkowski spacetime
which move near infinity under the influence of each other's Coulomb
field in the asymptotic past and future, property (1) does not
hold since $\tFR_{ab}$ fails to smoothly extend to ${\cal I}^-$ and
$\tFA_{ab}$ fails to smoothly extend to ${\cal I}^+$
\cite{willwalker,garfinkle}. Nevertheless, the ``pullback'' of
$\tFR_{ab}$ does continuously extend to ${\cal I}^-$; i.e., if we
contract $\tFR_{ab}$ into smooth two vector fields which, at ${\cal
I}^-$, are tangential to ${\cal I}^-$, then the resulting scalar field
continuously extends to ${\cal I}^-$. Since only these components of
$\tFR_{ab}$ are relevant for our energy flux calculations, it appears
that, even though result (1) fails, eq.~(\ref{pointEM}) will hold
nevertheless in this case. More generally, we expect
eq.~(\ref{pointEM}) to hold for particle orbits in curved spacetime
which come in from infinity in the asymptotic past in (nearly)
geodesic motion and emerge to infinity in the future also in (nearly)
geodesic motion, provided, of course, that the spacetime satisfies our
spacetime assumptions, the decay hypothesis, and the stationary
solution property.

If the spacetime $(M,g_{ab})$ is axisymmetric as well as stationary,
then essentially all of the results leading to Theorem 2.3 carry over
straightforwardly if we replace the timelike Killing field $t^a$ by
the axial Killing field $\phi^a$. The only exception is the analysis
of the ``advanced-retarded cross-term'' of eq.~(\ref{CTvanish}) above,
which now no longer automatically vanishes as a consequence of
eq.~(\ref{Fscri}). Nevertheless, under the additional assumptions that
point charge stationary solutions are axisymmetric on $\scri^+$ and
$\scri^-$ and that the vector potential of a source-free solution can
be chosen to vanish as one approaches timelike infinity along
$\scri^+$ and $\scri^-$, this cross-term can be seen to vanish by the
following argument.

For definiteness, let us consider the integral over past null
infinity. In light of our assumption that the point particle source
$j_a$ differs from the stationary source $\jS_a$ only in a compact
region of $M$, on $\scri^-$ we can write $\FR_{ab}$ and $\FA_{ab}$ as
\begin{eqnarray}
\FR_{ab} &=& \FS_{ab}\\
\FA_{ab} &=& \FS_{ab} + F'_{ab},
\end{eqnarray}
where $\FS_{ab}$ is the retarded (= advanced) solution associated
with $\jS_a$. Then $F'_{ab}$ is source-free outside of a compact
region of $M$ and vanishes outside of the causal past of this
region. It follows from eq.~(\ref{Fscri}) that $T_{ab}[\FS,\FS]$
radiates no angular momentum through $\scri^-$, so we have
\begin{eqnarray}
2\int_{\scri^-} \phi^a T_{ab}[\FR,\FA] \, \ep_{cde}{}^b 
  &=& 2\int_{\scri^-} \phi^a T_{ab}[\FS,F'] \, \ep_{cde}{}^b \nonumber \\
  &=& \int_{\scri^-} \phi^a \left(\FS_{af} F'_b{}^f + F'_{af} \FS_b{}^f
                                 - \frac{1}{2} g_{ab} \FS_{fg} F'^{fg}\right)
                     \,\ep_{cde}{}^b.
\end{eqnarray}
Since $\phi^a$ lies in $\scri^-$, the last term clearly
vanishes. Writing out $F'_{ab}$ and $\FS_{ab}$ in terms of the
corresponding vector potentials, we arrive at
\begin{eqnarray}
2\int_{\scri^-} \phi^a T_{ab}[\FR,\FA] \, \ep_{cde}{}^b 
  &=& \int_{\scri^-} \left(\phi^a \nabla_{[a}\AS_{f]} F'_b{}^f
                         + \phi^a \nabla_{[a}A'_{f]} \FS_b{}^f \right)
      \ep_{cde}{}^b \nonumber\\
  &=& \int_{\scri^-} \nabla_f (\phi^a \AS_a) F'^{bf} \, \ep_{cdeb}
      - \int_{\scri^-} \lie_{\phi} \AS_f F'^{bf} \, \ep_{cdeb} \nonumber\\
   && \mbox{} + \int_{\scri^-} \nabla_f (\phi^a A'_a) (\FS)^{bf}
                \, \ep_{cdeb}
      - \int_{\scri^-} \lie_{\phi} A'_f (\FS)^{bf} \, \ep_{cdeb} 
\end{eqnarray}

The second integral is immediately seen to vanish under our assumption
that the vector potential of a stationary solution can be chosen to be
axisymmetric at $\scri^-$. Integrating by parts in the fourth integral gives us
\begin{equation}
\int_{\scri^-} \lie_{\phi} A'_f (\FS)^{bf} \, \ep_{cdeb}
    = - \int_{\scri^-} A'_f \lie_\phi (\FS)^{bf} \, \ep_{cdeb},
\end{equation}
which vanishes under the same assumption. (The ``boundary term'' makes
no contribution since the orbits of $\phi^a$ are closed). For the
first integral, since $F'^{ab}$ satisfies Maxwell's equations, we have
\begin{equation}
\int_{\scri^-} \nabla_f (\phi^a \AS_a) F'^{bf} \, \ep_{bcde}
  = 3 \int_{\scri^-} d\left[(\phi \cdot \AS)
                                        *{F'}\right],
\end{equation}
where the two-form $*F'$ is the Hodge dual, defined by $*F'_{ab} =
- (1/2) \epsilon_{ab}{}^{cd} F'_{cd}$. Using Stokes' theorem, we can
convert this integral to two boundary integrals, one near $i^0$ and
the other near $i^-$. Since the two-form $*{F'}$ is source-free
outside of a compact region of $M$, the integrand vanishes near $i^0$
by causality and vanishes near $i^-$ under the decay hypothesis.
Similarly, for the third integral we have
\begin{equation}
\int_{\scri^-} \nabla_f (\phi^a \cdot A'_a) (\FS)^{bf} \, \ep_{cdeb}
  = 3 \int_{\scri^-} d\left[(\phi \cdot A')
                                        *{\FS}\right].
\end{equation}
This term vanishes because the vector potential $A'_a$ can be chosen
to vanish near $i^0$ by causality and, by our assumption, can be
chosen to vanish near $i^-$.

By reversing the roles of $\FR_{ab}$ and $\FA_{ab}$ and repeating the
above arguments, we also show that the integral over $\scri^+$
vanishes. Thus, if $(M,g_{ab})$ is axisymmetric as well as stationary
and the additional conditions given above are met, then under the
hypotheses of Theorem 2.3, eq.~(\ref{pointEM}) continues to hold if we
replace $t^a$ by $\phi^a$.

Finally, we note that our results can be generalized to allow for the
presence of a black hole and white hole whose horizons, ${\cal H}^+$
and ${\cal H}^-$, intersect on a compact 2-surface, $S$. We again
require $(M, g_{ab})$ be a globally hyperbolic, stationary spacetime
that is asymptotically flat at null and spatial infinity. However, we
now require that there exist a spacelike hypersurface, $\Sigma$, with
boundary, $S$, of the form $\Sigma = \Sigma_{\rm end} \cup \Sigma'$
with $\Sigma'$ compact, such that $\Sigma \setminus S$ is a Cauchy
surface for the domain of outer communications. In this case, we
deform $\Sigma$ to the future in small neighborhoods of both $i^0$ and
$S$ so that the deformed surface, $\tls^+$, intersects both $\scri^+$
and the future event horizon ${\cal H}^+$ in a cross-section. We
construct $\tls^-$ similarly.  With these changes, our previous
analysis for the case of no black or white holes can now be
repeated\footnote{The only exception is that for a rotating black
hole, the vanishing of the advanced-retarded cross-terms does not
follow from the horizon analog of eq.~(\ref{Fscri}) but instead must
be shown by the arguments similar to those used above for the case of
angular momentum. However, as we shall note shortly, the validity of
the stationary solution hypothesis appears to be limited to
non-rotating black holes in any case.} wherein in all of our
assumptions, arguments, and formulas, we replace ${\cal I}^+$ by
${\cal I}^+ \cup {\cal H}^+$ and ${\cal I}^-$ by ${\cal I}^- \cup
{\cal H}^-$. In
particular, Theorem 2.3 holds for the above class of spacetimes
containing black holes and white holes provided only that we modify
eq.~(\ref{pointEM}) so as to include contributions from the energy
flux through ${\cal H}^+$ and ${\cal H}^-$ in exact parallel with the
contributions from ${\cal I}^+$ and ${\cal I}^-$.

However, it would appear that some significant differences do occur in
the classes of spacetimes for which the decay hypothesis and
stationary solution property hold. Recall that when no black hole or
white hole is present, we conjectured that the decay hypothesis would
hold if and only if no ergoregion is present. However, for a rotating
black hole---i.e., if $t^a$ is not normal to ${\cal H}^+$ and ${\cal
H}^-$---the ``ergoregion instability'' argument~\cite{friedman} no
longer applies, since negative energy can be radiated into the black
hole. Thus, it seems plausible that the decay hypothesis will hold for
many rotating black holes---in particular, for Kerr black
holes---despite the presence of ergoregions. On the other hand, for
the case of a rotating black hole, it does not seem plausible that the
stationary solution property holds (where, with the modifications
described above, this property now includes the requirement that the
Maxwell field extend continuously to ${\cal H}^+$ and ${\cal
H}^-$). Specifically, for the retarded solution for a
non-axisymmetric, stationary source, there should be a steady,
nonvanishing flux of angular momentum into the black
hole. Consequently, $\FR_{ab}$ should not be continuously extendible
to ${\cal H}^-$. Similarly, $\FA_{ab}$ should fail to be continuously
extendible to ${\cal H}^+$.  However, it seems plausible that the
stationary solution property always holds for a non-rotating black
hole\footnote{In particular, in the static case, the equality of
$\FR_{ab}$ and $\FA_{ab}$ for static sources throughout the domain of
outer communications follows from the time reflection symmetry. Since
$\FR_{ab}$ is regular on ${\cal H}^+$ and $\FA_{ab}$ is regular on
${\cal H}^-$ it follows that both must have continuous limits to
${\cal H}^+ \cup {\cal H}^-$, in accord with the stationary solution
property. Note, however, that $\FR_{ab}$ will be discontinuous across
${\cal H}^-$ and $\FA_{ab}$ will be discontinuous across ${\cal H}^+$ in
a manner similar to the behavior of the retarded and advanced
solutions for a uniformly accelerating charge in Minkowski spacetime
\cite{boulware}.}. Thus, it appears likely that the hypotheses of
Theorem 2.3 hold precisely for the case of non-rotating black holes
that contain no ergoregion. Of course, the conclusion of that theorem
may still hold under more general hypotheses.

Finally, in the axisymmetric case, we may consider conservation of
angular momentum in spacetimes containing a non-rotating black hole
and white hole. As in the case where no black hole or white hole is
present, all of our arguments carry through straightforwardly when
$t^a$ is replaced by $\phi^a$ except for the vanishing of the
advanced-retarded cross term. However, the presence of a black and
white hole now produces a significant difference because, even in
Schwarzschild spacetime, the field of a stationary point charge fails
to be axisymmetric on the horizon. Thus, the symmetry argument used
above to obtain the vanishing of the advanced-retarded cross terms at
${\cal I}^+$ and ${\cal I}^-$ is inapplicable to ${\cal H}^+$ and
${\cal H}^-$. We have not been able to show that these cross terms
vanish (or that the cross term contribution from ${\cal H}^+$ cancels
that from ${\cal H}^-$). It would be of interest to determine if any
violations of angular momentum conservation can occur in this case
and, if so, how large the violations could be.

\section{Gravitational case}

In this section, we shall give an analysis of conservation results in
the gravitational case in close parallel with our analysis of the
electromagnetic case given in the previous section. Although there are
many similarities between the gravitational and electromagnetic cases,
there also are some key differences as well as some additional
technical difficulties occurring in the gravitational case.
Consequently, with the exception of Theorem 3.1 below, we shall not
attempt to prove theorems based upon analogs of our Maxwell field and
decay hypothesis assumptions, but rather will merely sketch how it
should be possible to obtain similar results. In this sense, our
results for point particles in the gravitational case will be
considerably weaker than in the electromagnetic case.

To begin, we wish to analyze energy conservation for solutions to the
linearized Einstein equation on a spacetime, $(M, g_{ab})$, which
satisfies the spacetime assumptions stated at the beginning of the
previous section. However, unlike the electromagnetic case, in order
to have well defined linearized field equations, it is necessary here
that the background spacetime be a solution to Einstein's
equation. Although there should not be any difficulty in principle in
allowing the background spacetime to possess suitable matter fields,
in order to directly apply the results of~\cite{quinnwald} in the case
of point particles, we must restrict to the case where the spacetime
is vacuum at least in a neighborhood of the worldline of the
particle. This restriction would still allow us to consider, for
example, a fluid star solution in which the particle remains exterior
to the star, but the inclusion of fluid or other matter fields would
significantly complicate the analysis. Hence, for simplicity, we shall
restrict attention to the case where the background spacetime
satisfies $G_{ab} = 0$. Unfortunately, this restricts the background
spacetime to be Minkowski spacetime.\footnote{This follows immediately
from the positive energy theorem together with the fact that the
(Komar) mass of a vacuum spacetime satisfying the spacetime
assumptions stated at the beginning of Sec.~II is easily seen to
vanish.} More generally, when black holes are admitted, the stationary
black hole uniqueness theorems restrict the background spacetime to
being Kerr. However, we shall not make explicit use of any special
properties of the background spacetime in our analysis below, and we
believe that our analysis should generalize to cases where suitable
matter is present in the background spacetime.

On $(M, g_{ab})$, we initially wish to consider smooth solutions,
$\gamma_{ab}$, to the linearized Einstein equation with a smooth
source $T_{ab}$
\begin{equation}
G^{(1)}_{ab} [\gamma_{cd}] = 8 \pi T_{ab}
\label{lingrav}
\end{equation}
Our basic strategy here will be to repeat the analysis of the previous
section, replacing the electromagnetic energy current three-form $t^a
T_{ab} \epsilon_{cde}{}^b$ with the ``effective gravitational energy
current'' three-form $t^a \tau_{ab} \epsilon_{cde}{}^b \equiv -
\frac{1}{8 \pi} t^a G^{(2)}_{ab} \epsilon_{cde}{}^b$ constructed from
the second order Einstein tensor $G^{(2)}_{ab}$ (see~\cite{habisohn})
associated with the linearized solution. In doing so, we will rely on
the results of Habisohn~\cite{habisohn}, which show that the net flux
from this effective energy current at null infinity agrees with the
net Bondi flux to second order. However, in order to apply Habisohn's
results, we need to assume that $T_{ab}$ vanishes in a neighborhood,
$U$, of $\scri^+ \cup \scri^- \cup i^0$ that contains complete Killing
orbits and that, in $U$, $\gamma_{ab}$ satisfies both the weak and
strengthened fall-off conditions of~\cite{habisohn}. Here, the weak
fall-off conditions of~\cite{habisohn} require that $\gamma_{ab}$
vanish in a neighborhood of $i^0$ and satisfy the decay properties in
$M \cap U$ stated in~\cite{habisohn}. The strengthened fall-off
conditions require that, when transformed to the Geroch-Xanthopoulos
gauge~\cite{gx}, the unphysical metric perturbation
$\tilde{\gamma}_{ab} = \Omega^2 \gamma_{ab}$ (which smoothly extends
to $\scri^+$ and $\scri^-$ in this gauge) satisfies similar decay
properties in $\tilde{M} \cap U$, where $\tilde{M}$ denotes the
conformally completed spacetime. The requirement that $\gamma_{ab}$
vanish in a neighborhood of $i^0$ is much stronger than condition (3)
imposed in the electromagnetic case, and, indeed, is too strong to
enable us to use it directly for our purposes below. Undoubtedly,
Habisohn's conditions could be weakened considerably, but this would
require further analysis that we do not wish to undertake here.

Let $g_{ab} (\lambda) = g_{ab} + \lambda \gamma_{ab}$, where $G_{ab}
[g_{cd}] = 0$ and $\gamma_{ab}$ satisfies eq.~(\ref{lingrav}).  The
Bianchi identity to second order in $\lambda$ yields
\begin{equation}
\frac{1}{2}\frac{d^2}{d \lambda^2}
   \left[\nabla_a^{(\lambda)} G^{ab}(\lambda) \right]_{\lambda=0}
   = G^{(2)ab} [\gamma_{cd}] + \nabla^{(1)}_a G^{(1)ab}[\gamma_{cd}] = 0;
\label{bianchi1}
\end{equation}
i.e.,
\begin{equation}
\nabla_a G^{(2)ab}[\gamma_{cd}] = - {C^a}_{ad}[\gamma_{cd}] G^{(1)db}[\gamma_{cd}] - {C^b}_{ad}[\gamma_{cd}]
G^{(1)ad}[\gamma_{cd}]
\label{bianchi2}
\end{equation}
where
\begin{equation}
{C^a}_{bc}[\gamma_{cd}] = \frac{1}{2} g^{ad} [\nabla_b \gamma_{cd} + \nabla_c
\gamma_{bd} - \nabla_d \gamma_{bc}]
\label{C}
\end{equation}
and where $\nabla_a$ denotes the derivative operator associated with
$g_{ab}$. Since $\gamma_{ab}$ satisfies
the linearized Einstein equation (\ref{lingrav}), we find
\begin{equation}
\nabla_a G^{(2)ab} =  - 8 \pi [C^a{}_{ac} T^{cb} - C^b{}_{ac}
T^{ac}]
\label{consgrav}
\end{equation}

Equation (\ref{consgrav}) is a direct analog of eq.~(\ref{Tcons}) in
the electromagnetic case. Furthermore, although in this case the
physical ``effective energy current three-form''
\begin{equation}
t^a \tau_{ab} \epsilon_{cde}{}^b \equiv - \frac{1}{8 \pi} t^a
G^{(2)}_{ab} \epsilon_{cde}{}^b
\label{tau}
\end{equation}
need not continuously extend to null infinity, Habisohn
\cite{habisohn} has shown that if his weak and strengthened fall-off
conditions hold, its pullback to certain timelike 3-surfaces has a
continuous limit as these surfaces approach null
infinity. Consequently, in exact parallel with the derivation of
eq.~(\ref{smoothEM}), we obtain the following theorem:

\noindent {\bf Theorem 3.1} Let $(M, g_{ab})$ be a vacuum solution of
Einstein's equation satisfying the spacetime assumptions\footnote{As
previously noted, these assumptions actually imply that $(M, g_{ab})$
is Minkowski spacetime. However, as also previously noted, the theorem
can be generalized to allow for black holes (in which case the Kerr
family of solutions would be admissible) and, presumably, also could
be generalized to allow solutions to the nonvacuum Einstein equation
with suitable forms of matter.} of Sec.~II. Let $\gamma_{ab}$ be a
smooth solution to the linearized Einstein equation with smooth source
$T_{ab}$ such that $T_{ab}$ vanishes in a neighborhood, $U$, of
$\scri^+ \cup \scri^- \cup i^0$ that contains complete Killing orbits
and such that, in $U$, $\gamma_{ab}$ satisfies both the weak and
strengthened fall-off conditions of~\cite{habisohn}. Then,
\begin{eqnarray}
\lim_{t^+ \rightarrow +\infty} \int_{\tls^+(t^+)}
    \!\! t^a \tau_{ab} \epsilon_{cde}{}^b
  - \lim_{t^- \rightarrow -\infty} \int_{\tls^-(t^-)}
    \!\! t^a \tau_{ab} \epsilon_{cde}{}^b && \nonumber \\
  \mbox{} + \int_{\scri^+} \!\! t^a \tau_{ab} \epsilon_{cde}{}^b
  - \int_{\scri^-} \!\! t^a \tau_{ab} \epsilon_{cde}{}^b 
   &=& - \int_{M} \!\! t_a [{C^b}_{bc} T^{ca} + {C^a}_{bc} T^{bc}]
                       \epsilon_{defg}.
\label{smoothgrav}
\end{eqnarray}
provided that the integral over $M$ and the limits of the integrals
over $\tls^+(t^+)$ and $\tls^-(t^-)$ exist. Here, the integrals over
$\scri^+$ and $\scri^-$ are understood to mean the limits of the
integrals over suitable timelike surfaces in the physical spacetime
which approach $\scri^+$ and $\scri^-$; these limits were proven by
Habisohn~\cite{habisohn} to exist.

\medskip

The second pair of terms in eq.~(\ref{smoothgrav}) was proven by
Habisohn~\cite{habisohn} to agree, to second order in $\gamma_{ab}$,
with the net Bondi energy radiated to infinity. In analogy with the
electromagnetic case, it would be natural to attempt to interpret the
first pair of terms as representing the difference between the initial
and final gravitational energy of the matter distribution, and,
similarly, to interpret the right side as representing the
work done by the matter on the gravitational field. However,
unlike the electromagnetic case, these terms are, in general, gauge
dependent. Thus, unless some further restrictions and/or gauge
conditions are imposed, such an interpretation would not, in general,
be meaningful. Nevertheless, under the hypotheses of the theorem, the
second pair of terms is gauge independent---provided, of course that
one stays in a gauge where the weak fall-off conditions hold.

In order to proceed further, in analogy with the analysis given for
the electromagnetic case, we need to restrict consideration to
situations where the first pair of terms in eq.~(\ref{smoothgrav})
cancel. As in the electromagnetic case, we consider the case where
$T_{ab}$ differs from a stationary matter distribution only in a
compact region of spacetime. However, in order that $\gamma_{ab}$
satisfy the decay properties needed for cancellation of the first pair
of terms in eq.~(\ref{smoothgrav}), it also is necessary to impose
gauge conditions on $\gamma_{ab}$. Note that if a choice of gauge
compatible with the Habisohn fall-off conditions~\cite{habisohn} has
been made so that the first pair of terms in eq.~(\ref{smoothgrav})
cancel, the right side of that equation automatically must be gauge
invariant with respect to any remaining restricted gauge
freedom. Therefore, in that case it would seem appropriate to
interpret the right side of that equation as corresponding to the work
done by the matter on the gravitational perturbation, $\gamma_{ab}$,
from the ``viewpoint'' of the background spacetime, $(M, g_{ab})$.

A choice of gauge which appears to be suitable for the above purpose
is the Lorentz gauge condition,
\begin{equation}
\nabla^a \bar{\gamma}_{ab} = 0
\label{lorentz}
\end{equation}
where
\begin{equation}
\bar{\gamma}_{ab} = (\gamma_{ab} - \frac{1}{2} \gamma g_{ab})
\label{gammabar}
\end{equation}
In the Lorentz gauge, $\bar{\gamma}_{ab}$ satisfies a wave equation,
\begin{equation}
\nabla^c \nabla_c \bar{\gamma}_{ab} - 2 {{R^c}_{ab}}^d
\bar{\gamma}_{cd} = - 16 \pi T_{ab}
\label{wave}
\end{equation}
and it seems highly plausible that, in a wide class of spacetimes,
suitable decay properties at late and early times will hold for source
free solutions with suitable fall-off near $i^0$. In particular, it
seems plausible that for source free solutions with suitable fall-off
near $i^0$, in this gauge the supremum over $\tls^+(t^+)$ of the
components of $\gamma_{ab}$ and its first two derivatives will go to
zero as $t^+ \rightarrow \infty$, and that similar decay properties
will hold as $t^- \rightarrow - \infty$. It also seems plausible that,
for such solutions in this gauge, the stronger fall-off conditions of
\cite{habisohn} near $\scri^+$ and $\scri^-$ will hold as $t^+
\rightarrow \infty$, and $t^- \rightarrow - \infty$. Finally, it
appears highly plausible that, in the Lorentz gauge, an analog of the
stationary solution property (see Sec.~II) will hold. Consequently,
a direct analog of theorem 2.2 should hold in the gravitational case
when we impose the Lorentz gauge condition. However, we shall not
attempt to state and prove such a theorem here, since it would require
considerable further analysis to make a judicious choice of
mathematically precise formulations of the gravitational versions of
the decay hypothesis and the other relevant assumptions needed for the
proof of such a theorem.

We may now attempt ask in the gravitational case whether, for a point
particle source whose motion differs from a Killing orbit for only a
finite time interval, the analog of eq.~(\ref{pointEM}) holds, i.e.,
whether
\begin{equation}
\int_{\scri^+} \!\! t^a \tau_{ab} \epsilon^b{}_{cde}
  - \int_{\scri^-} \!\! t^a \tau_{ab} \epsilon^b{}_{cde}
  =  \int t^a f_a \, d\tau.
\label{pointgrav}
\end{equation}
where the integrals over $\scri^+$ and $\scri^-$ have the same meaning
as in theorem 3.1, and where $f^a$ now denotes the gravitational
self-force~\cite{minoetal,quinnwald},
\begin{eqnarray}
f^a &=& m\left(\frac{1}{2} \nabla^a \gin_{bc}
	- \nabla_b (\gin)_c{}^a \right) u^b u^c 
      	- m^2 \left(\frac{11}{3} \dot{a}^a + \frac{1}{3} a^2 u^a \right)
\nonumber \\
&& + m^2 u^b u^c \int_{-\infty}^{\tau} 
	\left(\frac{1}{2} \nabla^a \GR_{bca'b'} - \nabla_b (\GR)_c{}^a{}_{a'b'}
 	      \right) u^{a'} u^{b'}
        \,d\tau'
\label{fgrav}
\end{eqnarray}
Unfortunately, however, this
question as posed has only a trivial domain of applicability: As
emphasized in~\cite{quinnwald}, in order for a solution to the
linearized Einstein equation with a point particle source to exist, it
is necessary that the particle move on a geodesic. Therefore, a
point particle cannot follow a Killing orbit for a finite time
interval unless that Killing orbit is a geodesic, in which case the
particle's motion can never deviate from the Killing orbit. In other
words, the constraints on the motion of a particle imposed by the
linearized Einstein equation preclude the possibility of having a
particle which is nonstationary for only a finite time interval. Note
that this contrasts sharply with the situation in the electromagnetic
case, where Maxwell's equations impose no constraints on the motion of
a charged particle.

Nevertheless, it is worth noting that it is not manifestly
inconsistent to have general distributional sources containing point
masses which are stationary outside of a compact spacetime region. In
other words, we may have a system of point masses connected by devices
such as rods, springs, or strings such that the point masses are
initially and finally in the same stationary position, but are set
into motion at intermediate times. Under decay assumptions of the type
expected to hold in the Lorentz gauge as described above, it then
should be possible to prove a direct analog of Proposition 2.1. Namely, if
$T_{ab}$ now denotes a distributional stress-energy tensor which is
stationary outside of a compact spacetime region, and if $\gamma_{ab}$
denotes a (distributional) solution to the linearized Einstein
equation with source $T_{ab}$ such that $\gamma_{ab}$ is smooth near
infinity and satisfies suitable decay properties near $i^0$, then it
should follow that
\begin{equation}
\int_{\scri^+} \!\! t^a \tau_{ab} \epsilon_{cde}{}^b
  - \int_{\scri^-} \!\! t^a \tau_{ab} \epsilon_{cde}{}^b 
   = - \int_{M} \!\! t_a [(\Cfree)^b{}_{bc} T^{ca} + (\Cfree)^a{}_{bc} T^{bc}]
                       \epsilon_{defg}.
\label{prop2}
\end{equation}
where $(\Cfree)^a{}_{bc}$ is given by eq.~(\ref{C}), with $\gamma_{ab}$
replaced by $\gfree_{ab}$, where
\begin{equation}
\gfree_{ab} \equiv \gamma_{ab} - \gbar_{ab}
\end{equation}
with
\begin{equation}
\gbar_{ab} = \frac{1}{2} [\gA_{ab} + \gR_{ab}].
\end{equation}
Note that in eq.~(\ref{prop2}) it is, in general, necessary to take
into account the effect of the rods and other devices both directly
with regard to their contribution to $T_{ab}$ and indirectly with
regard to their contribution to the gravitational perturbation
$\gamma_{ab}$.

Although we cannot even pose a meaningful question for a system
composed purely of point masses which are stationary in the past and
future, we can consider energy conservation for a freely falling point
mass which follows a trajectory which starts at infinity and returns
to infinity. In a general, curved spacetime\footnote{As already
mentioned, our above spacetime assumptions restrict us to flat
spacetime, but the remarks here are relevant for generalizations to
allow for the presence of black holes as well as
generalizations to non-vacuum cases.} such a particle of mass $m$ will
radiate energy of order $m^2$ to infinity, and one may ask if this
agrees, to lowest order, with minus the net work done on the particle by the
gravitational self-force. On account of this gravitational self-force,
the particle will fail to move on a geodesic, resulting in
self-consistency issues with regard to treating its motion that were
discussed in~\cite{quinnwald}. However, in order to calculate the work
done by the gravitational self-force to order $m^2$, the deviations
from geodesic motion produced by the gravitational self-force can be
neglected. Thus, we may ask whether eq.~(\ref{pointgrav}) holds for a
particle moving on a geodesic.

In parallel with our discussion in the electromagnetic case (see the
paragraph below eq.~(\ref{Ueff}) of Sec.~II), if suitable decay
properties at late and early times hold for source free solutions, if
the advanced and retarded solutions for the particle moving on the
given geodesic are suitably well behaved, and if the
``advanced-retarded cross-term'' makes no contribution to the flux of
$t^a \tau_{ab} \epsilon_{cde}{}^b$ through $\scri^+$ and $\scri^-$,
then eq.~(\ref{prop2}) should hold, i.e.,
\begin{equation}
\int_{\scri^+} \!\! t^a \tau_{ab} \epsilon_{cde}{}^b
  - \int_{\scri^-} \!\! t^a \tau_{ab} \epsilon_{cde}{}^b
  =  \int t^a \fg_a \, d\tau.
\end{equation}
where
\begin{eqnarray}
\fg^a &=& m\left(\frac{1}{2} \nabla^a \gin_{bc}
	- \nabla_b (\gin)_c{}^a \right) u^b u^c
\nonumber \\
&& + \frac{1}{2} m^2 u^b u^c \int_{-\infty}^{\tau} 
	\left(\frac{1}{2} \nabla^a \GR_{bca'b'} - \nabla_b (\GR)_c{}^a{}_{a'b'}
 	\right) u^{a'} u^{b'}
\,d\tau'
\nonumber \\
&& - \frac{1}{2} m^2 u^b u^c \int_{\tau}^{+\infty} 
	\left(\frac{1}{2} \nabla^a \GA_{bca'b'} - \nabla_b (\GA)_c{}^a{}_{a'b'}
 	      \right) u^{a'} u^{b'}
\,d\tau'
\label{fgravgaltsov}
\end{eqnarray}
However, the difference between $f^a$ and $\fg^a$ is given by
\begin{equation}
f^a - \fg^a = m^2 u^b u^c \int_{-\infty}^{+\infty} 
	\left(\frac{1}{2} \nabla^a \Gbar_{bca'b'}
              - \nabla_b \Gbar_c{}^a{}_{a'b'}\right)
	u^{a'} u^{b'}
        \,d\tau'
\end{equation}
Our task is to show that, just as in the electromagnetic
case, this difference between the two forces does no net work over the
world line of the particle. 

In analogy with eq.~(\ref{workdiff}), we have
\begin{eqnarray}
\int_{\tau^-}^{\tau^+} t^a (f_a-\fg_a) \, d\tau
  &=& m^2 \int_{\tau^-}^{\tau^+} t^a \left[
            \int_{-\infty}^{+\infty} 
	    \left(\frac{1}{2} \nabla_a \Gbar_{bcb'c'}
                  - \nabla_b \Gbar_{cab'c'}
            \right) u^{b'} u^{c'} \,d\tau'
           \right] u^b u^c \, d\tau
\nonumber \\
  &=& m^2 \int_{\tau^-}^{\tau^+} \int_{-\infty}^{+\infty} 
	      \left(\frac{1}{2} \lie_t \Gbar_{bcb'c'} 
                    - \Gbar_{acb'c'} \nabla_b t^a
                    - t^a \nabla_b \Gbar_{cab'c'} \right)
	      u^b u^c u^{b'} u^{c'} \,d\tau' \, d\tau
\nonumber \\
  &=& \frac{m^2}{2} \int_{\tau^-}^{\tau^+} \int_{-\infty}^{+\infty} 
	      \lie_t \Gbar_{bcb'c'} 
              u^b u^c u^{b'} u^{c'} \,d\tau' \, d\tau
\nonumber \\ \mbox{}
  && - m^2 \int_{\tau^-}^{\tau^+} \int_{-\infty}^{+\infty} 
                 u^b \nabla_b (t^a u^c \Gbar_{acb'c'})
	         u^{b'} u^{c'} \,d\tau' \, d\tau
\label{gravworkdiff}
\end{eqnarray}
By arguments similar to those which produced eqs.~(\ref{Vdiff})
and~(\ref{Lieterm}), both of these integrals can be reduced to
expressions involving ``tail contributions'' on the worldline of the
particle for $\tau \geq \tau^+$ and $\tau \leq \tau^-$.  Since here we
restrict consideration to geodesic particle trajectories which start
at infinity and return to infinity in asymptotically flat spacetimes,
these tail contributions should vanish in the limit $\tau^\pm
\rightarrow \pm \infty$. Thus, eq.~(\ref{pointgrav}) should hold in
this case, although the reader surely will have noted the numerous gaps in
our arguments that would have to be filled in order to convert our arguments
into a theorem.

Finally, we comment that, in parallel with the electromagnetic case,
our arguments should generalize to yield a conservation of angular
momentum result in the case of a spacetime that is axisymmetric as
well as stationary. Similarly, our arguments should generalize
straightforwardly to the case where black holes and white holes are
present provided that the fluxes through the black hole and white hole
horizons are included in the manner described near the end of Sec.~II.

\section*{Acknowledgements}

This research was supported in part by NSF grant PHY 95-14726 to the
University of Chicago.

\appendix

\section{Massless Klein-Gordon scalar case}

In this appendix, we outline the derivation of conservation results
for the case of a massless Klein-Gordon field, satisfying
\begin{equation}
\nabla^a \nabla_a \phi = - j.
\label{scalfldeq}
\end{equation}
To a large extent, the analysis of energy and angular momentum
conservation for this case closely parallels the electromagnetic case
analyzed in Sec.~II, although there are some complications arising
from the lack of conformal invariance of the theory and the important
difference that the Klein-Gordon force need not be orthogonal to the
four-velocity.

The stress-energy tensor of the massless Klein-Gordon scalar field is
given by
\begin{equation}
T_{ab} = \nabla_a \phi \nabla_b \phi
         - \frac{1}{2} g_{ab} g^{cd} \nabla_c \phi \nabla_d \phi
\end{equation}
By virtue of eq.~(\ref{scalfldeq}), it satisfies
\begin{equation}
\nabla^b T_{ab} = - j \nabla_a \phi.
\label{scastresscons}
\end{equation}
In precise analogy with the electromagnetic case, we define the energy
three-current three-form by
\begin{equation}
J_{cde} = t^a T_{ab} \ep_{cde}{}^b.
\end{equation}
It follows that
\begin{equation}
(dJ)_{cdef} = j t^a \nabla_a \phi \ep_{cdef}.
\label{dJphi}
\end{equation}

The task of determining falloff conditions on $\phi$ at null infinity
is less trivial here as compared with the Maxwell case since the field
equation~(\ref{scalfldeq}) is not conformally invariant. However, the
standard conformally invariant modification of
eq.~(\ref{scalfldeq}), namely $\nabla^a \nabla_a \phi - \frac{1}{6} R
\phi= - j$, is equivalent to eq.~(\ref{scalfldeq}) in vacuum
spacetimes. This suggests that we define an ``unphysical Klein-Gordon
scalar field'', $\phit$, by using the same conformal weight as in the
conformally invariant case, i.e.,
\begin{equation}
\phit = \Om^{-1} \phi 
\end{equation}
and require that $\phit$ extend smoothly to $\scri^+$ and
$\scri^-$. We adopt this condition as the analog of the Maxwell field
assumption (2) of Sec.~II.

In terms of unphysical variables, the physical energy
three-current takes the form
\begin{eqnarray}
J_{cde} &=& t^a
            \left[\nabla_a \phi \nabla_b \phi
                  - \frac{1}{2} g_{ab} g^{cd}
                    \nabla_c \phi \nabla_d \phi
            \right] 
            g^{bf} \ep_{cdef} \nonumber \\
  &=& t^a
      \left[\nt_a\phit \nt_b\phit
            + 2 \Om^{-1} \phit \nt_{(a} \phit \nt_{b)}\Om
            + \Om^{-2} \phit^2 \nabla_a \Om \nabla_b \Om
\right. \nonumber \\ && \phantom{\frac{1}{4\pi} t^a\left[\right.}\left.
            - \frac{1}{2} \gt_{ab} \gt^{cd}
              (\nt_c\phit \nt_d\phit
            + 2 \Om^{-1} \phit \nt_c \phit \nt_d\Om
            + \Om^{-2} \phit^2 \nt_c \Om \nt_d \Om)
       \right] 
       \gt^{bf} \ept_{cdef}
\end{eqnarray}
where $\nt_a$ denotes the derivative operator associated with the
unphysical metric $\gt_{ab}$ (although it should be noted that in this
formula, only derivatives of scalar fields are taken, so there
actually is no distinction between $\nt_a$ and $\nabla_a$). Since $n_a
\equiv \nabla_a \Omega \neq 0$ on $\scri \equiv \scri^+ \cup \scri^-$
we see that $J_{cde}$ fails to extend continuously to $\scri$ even
when $\phit$ extends smoothly to $\scri$. However, the pull-back,
$\overline{J}_{cde}$, of $J_{cde}$ to a surface of constant $\Omega$ is
given by
\begin{equation}
\overline{J}_{cde}  = - \left[
      t^a \nt_a \phit n^b \nt_b\phit
            + f \phit t^a \nt_a \phit
            + \frac{1}{2} f h \phit^2  
            - \frac{1}{2} \Om h \gt^{cd} \nt_c\phit \nt_d\phit
       \right] 
       {}^{(3)}\overline{\ept}_{cde},
\label{Jbar}
\end{equation}
where the functions $f \equiv \Om^{-1} \gt^{ab} n_a n_b$ and $h \equiv
\Om^{-1} t^a n_a$ extend smoothly to $\scri$, and ${}^{(3)}\ept_{abc}$
is defined (up to a three-form with vanishing pullback) by
\begin{equation}
\ept_{abcd} = n_{[a} {}^{(3)}\ept_{bcd]}
\end{equation}
Consequently, although $J_{cde}$ fails to extend continuously to
$\scri$, it can be seen by inspection of eq.~(\ref{Jbar}) that
$\overline{J}_{cde}$ extends smoothly to $\scri$. Equivalently, if we
choose any three smooth vector fields in the unphysical spacetime
which are everywhere tangent to surfaces of constant $\Omega$ and we
contract these vector fields with the free indices of $J_{cde}$, then
the resulting function will extend smoothly to $\scri$.

However, it is not difficult to verify that, in general, the limit of
$\overline{J}_{cde}$ to $\scri$ will depend upon the choice of
conformal factor, $\Omega$, used to define the surfaces to which
$J_{cde}$ is pulled back. Nevertheless, we argue now that the {\it
integral} over $\scri^+$ or $\scri^-$ of $\overline{J}_{cde}$ cannot
depend upon the choice of $\Omega$, provided only that the scalar
source, $j$, vanishes in a neighborhood of $\scri \cup i^0$ and that
$\phit$ has suitable fall-off properties at $i^0$ and at timelike
infinity. Namely, let $\Omega$ and $\Omega'$ be two choices of
conformal factor. Integrate $(dJ)_{cdef}$ over the region bounded by a
surface of constant $\Omega$ and a surface of constant $\Omega'$, with
suitable ``endcaps'' inserted near $i^0$ and future timelike
infinity. Now apply Stokes' theorem---using eq.~(\ref{dJphi}), the
vanishing of $j$ near $\scri$, and the decay of $\phi$ near $i^0$ and
timelike infinity---to conclude that, in the limit as one approaches
$\scri^+$, the integral of $J_{cde}$ over a surface of constant
$\Omega$ equals the integral of $J_{cde}$ over a surface of constant
$\Omega'$. From this it follows that the integral of
$\overline{J}_{cde}$ over $\scri^+$ is independent of the choice of
$\Omega$. Similar results apply, of course, to integrals over
$\scri^-$. Thus, if $\phit$ is smooth at $\scri$ and satisfies
suitable fall-off properties at $i^0$ and at timelike infinity, and if
$j$ vanishes in a neighborhood of $\scri \cup i^0$, then the total
energy flux through $\scri^+$ and $\scri^-$ is well defined.

For a spacetime satisfying the assumptions listed at the beginning of
Sec.~II, we integrate eq.~(\ref{dJphi}) over the compact region with
``endcaps'' given by $\tls^+(t^+)$ and $\tls^-(t^-)$ but now bounded
on the ``sides'' by a surface of constant, nonzero $\Omega$ (rather
than by $\scri \cup i^0$). We apply Stokes' theorem, then take the
limit as $t^\pm \rightarrow \pm \infty$, and, finally take the limit
as $\Omega \rightarrow 0$. The result is the analog of Theorem 2.1,
namely
\begin{eqnarray}
\lim_{t^+ \rightarrow +\infty} \int_{\tls^+(t^+)}
    \!\! t^a T_{ab} \epsilon_{cde}{}^b
  - \lim_{t^- \rightarrow -\infty} \int_{\tls^-(t^-)}
    \!\! t^a T_{ab} \epsilon_{cde}{}^b && \nonumber \\
  \mbox{} + \int_{\scri^+} \!\! t^a T_{ab} \epsilon_{cde}{}^b
  - \int_{\scri^-} \!\! t^a T_{ab} \epsilon_{cde}{}^b 
   &=& \int_{M} \!\! j t^a \nabla_a \phi \epsilon_{cdef},
\label{smoothscal}
\end{eqnarray}
where the integrals over $\scri^+$, $\scri^-$, $\tls^+(t^+)$, and
$\tls^-(t^-)$ are defined by the limiting procedure described above.
Assuming suitable analogs of the decay hypothesis and stationary
solution property of Sec.~II, we may then obtain an analog of
Theorem 2.2.

The analog of the Lorentz force on a particle of scalar charge $q$ in
an external Klein-Gordon field $\phi$ is
\begin{equation}
f_a = q \nabla_a \phi
\label{testsca}
\end{equation}
(This equation can be derived by integrating eq.~(\ref{scastresscons})
over a small body and neglecting the self-field of the body; the
Lorentz force in electromagnetism can be similarly derived by
integration of eq.~(\ref{Tcons}).) It should be noted that, in sharp
constrast with the electromagnetic case, the scalar force
(\ref{testsca}) fails, in general, to be perpendicular to the
4-velocity, $u^a$, of the particle. Consequently, the rest mass, $m$,
of the particle will vary with time, i.e., in general, the particle
will necessarily gain or lose rest mass as a result of its
interactions with the scalar field. Indeed, we have
\begin{equation}
\frac{dm}{d \tau} = - u^a f_a = - q u^a \nabla_a \phi
\end{equation}
Thus, the 4-momentum of a particle of scalar charge $q$ in an
external Klein-Gordon field $\phi$ is given by\footnote{It also should
be noted that---in sharp contrast with the electromagnetic
case---conservation of charge is not required for consistency of the
Klein-Gordon equation, so there is no obstacle to allowing $q$ to vary
with time as well. However, we shall assume throughout this Appendix
that $q$ is constant.}
\begin{equation}
p_a = (m_0 - q \phi) u_a
\label{pdef}
\end{equation}
where $m_0$ is a constant.

When self-field effects are taken into account, the formula for the
total force on a test particle (analogous to the DeWitt Brehme formula
in the electromagnetic case) is~\cite{quinnwiseman}
\begin{eqnarray}
f_a &=& q \nabla_a \phiin + q^2 \left(
        \frac{1}{3} \dot{a}_a
        - \frac{1}{3} a^2 u_a
        + \frac{1}{6} R_{ab} u^b
        + \frac{1}{6} u_a R_{bc} u^b u^c
        - \frac{1}{12} R u_a
        \right) \nonumber\\
&& \mbox{} + q^2 \int_{-\infty}^{\tau} \!\! \nabla_a \GR \,d\tau'
\label{scalarforce}
\end{eqnarray}
We wish to establish the analog of Theorem 2.3 for the scalar case,
i.e., we wish to show that if $z(\tau)$ is a timelike curve which
differs from an orbit, $z_0 (\tau)$, of the stationary Killing field
$t^a$ only over a finite interval, then
\begin{equation}
\int_{\scri^+} \!\! t^a T_{ab} \epsilon_{cde}{}^b
  - \int_{\scri^-} \!\! t^a T_{ab} \epsilon_{cde}{}^b
  =   \int t^a f_a \, d\tau.
\end{equation}
As in the electromagnetic case, we proceed by first establishing the
analog of Proposition 2.1, namely that
\begin{equation}
\int_{\scri^+} \!\! t^a T_{ab} \epsilon_{cde}{}^b
  - \int_{\scri^-} \!\! t^a T_{ab} \epsilon_{cde}{}^b
  =  \int t^a \fg_a \, d\tau
\label{scalpropeq}
\end{equation}
where
\begin{equation}
\fg_a = q \nabla_a \phifree
\label{scalfg}
\end{equation}
The only significant difference from the electromagnetic case
occurring here is in the analysis of the vanishing of the
``advanced-retarded cross-term'' in the integrals representing the
energy radiated through $\scri$. This cross-term vanished trivially in
the electromagnetic case, but here is given by a nontrivial
expression. However, we can simplify this expression by choosing
$\Omega$ so that $\lie_t \Om = t^a n_a = 0$. For the integral of the
cross term over $\scri^+$, we have
\begin{eqnarray}
2 \int_{\scri^+} \!\!\overline{J}_{cde}[\phiA,\phiR]
  &=& - \int_{\scri^+}\!\!
      \left[
      t^a \nt_a \phiAt n^b \nt_b\phiRt
      + t^a \nt_a \phiRt n^b \nt_b\phiAt \right.
\nonumber \\ &&
\phantom{- \int_{\scri^+}\!\!
         \left[ \right.} \left.
            \mbox{} + f \phiAt t^a \nt_a \phiRt
            + f \phiRt t^a \nt_a \phiAt
       \right] 
       {}^{(3)}\overline{\ept}_{cde} \nonumber \\
  &=& - \int_{\scri^+}\!\!
      \left[
      t^a \nt_a \phiRt n^b \nt_b\phiAt
            + f \phiAt t^a \nt_a \phiRt
       \right] 
       {}^{(3)}\overline{\ept}_{cde} \nonumber \\
  &=& - \int_{\scri^+}\!\!
      \lie_t \left[
      \phiRt n^b \nt_b\phiAt
            + f \phiAt \phiRt
       \right] 
       {}^{(3)}\overline{\ept}_{cde} \nonumber \\
  &=& 0.
\end{eqnarray}
Here, we have used the fact that $\phiAt$ is a stationary solution
near $\scri^+$ so that $\lie_t \phiAt = t^a \nt_a \phiAt = 0$.  In the
last line, we have assumed that suitable falloff conditions on
$\phiRt$ near $i^0$ and timelike infinity lead to the vanishing of the
boundary terms.  Therefore, we see that the ``advanced-retarded
cross-term'' makes no contribution to the energy radiated through
$\scri$. The remainder of the derivation of eq.~(\ref{scalpropeq})
follows in close parallel with the electromagnetic case.

To compare the work done by $f_a$ with that done by $\fg^a$, we note that
$\fg_a$ is given by
\begin{eqnarray}
\fg_a &=& q \nabla_a \phiin + q^2 \left(
        \frac{1}{3} \dot{a}_a
        - \frac{1}{3} a^2 u_a
        + \frac{1}{6} R_{ab} u^b
        + \frac{1}{6} u_a R_{bc} u^b u^c
        - \frac{1}{12} R u_a
        \right) \nonumber \\
&& \mbox{} + \frac{q^2}{2} \int_{-\infty}^{\tau} \!\! \nabla_a \GR \,d\tau'
           - \frac{q^2}{2} \int_{\tau}^{+\infty} \!\! \nabla_a \GA \,d\tau'.
\end{eqnarray}
Hence, the difference between the total work done by $f_a$ and the
total work done by $\fg_a$ between $\tau=\tau^-$ and $\tau=\tau^+$ is
given by
\begin{eqnarray}
\int_{\tau^-}^{\tau^+} \!\! t^a(f_a - \fg_a)\, d\tau
  &=& q^2 \int_{\tau^-}^{\tau^+} \!\! t^a\left[\int_{-\infty}^{+\infty} \!\!
                             \nabla_a \Gbar \,d\tau'\right]\, d\tau
\nonumber \\
  &=& q^2 \int_{\tau^-}^{\tau^+} \!\! \int_{\tau^+}^{+\infty} \!\!
               t^a \nabla_a \Gbar \,d\tau' \, d\tau
    + q^2 \int_{\tau^-}^{\tau^+}\!\! \int_{-\infty}^{\tau^-} \!\!
               t^a \nabla_a \Gbar \,d\tau' \, d\tau
\nonumber \\
  &=& - q^2 \int_{\tau^-}^{\tau^+} \!\! \int_{\tau^+}^{+\infty} \!\!
               t^{a'} \nabla_{a'} \Gbar \,d\tau' \, d\tau
      - q^2 \int_{\tau^-}^{\tau^+}\!\! \int_{-\infty}^{\tau^-} \!\!
               t^{a'} \nabla_{a'} \Gbar \,d\tau' \, d\tau
\end{eqnarray}
where the scalar analogs of eqs.~(\ref{sym2}) and (\ref{symint}) were
used.  If we now take the limit $\tau^\pm \rightarrow \pm \infty$,
then $t^{a'} = \chi({\bf x}^+) u^{a'}$ in the $\tau'$ region of
integration of the first integral, where ${\bf x}^+$ denotes the final
spatial position of the particle and $\chi \equiv (-t^a
t_a)^{1/2}$. Thus, the integrand is a total derivative in
$\tau'$. Assuming that $\Gbar$ falls off so that
$\Gbar(z(\tau),z(\pm\infty)) = 0$, we therefore have
\begin{eqnarray}
\lim_{\tau^\pm \rightarrow \pm\infty}
  \int_{\tau^-}^{\tau^+} \!\! t^a(f_a - \fg_a)\, d\tau
  &=& \lim_{\tau^\pm \rightarrow \pm\infty} q^2 \int_{\tau^-}^{\tau^+} \!\! 
               [\chi({\bf x}^+) \Gbar(z(\tau),z(\tau^+)) -\chi({\bf x}^-) \Gbar(z(\tau),z(\tau^-))]
               \, d\tau
\nonumber \\
  &=& \lim_{\tau^\pm \rightarrow \pm\infty} \frac{q^2}{2}
      \int_{\tau^-}^{\tau^+} \!\! 
               [\chi({\bf x}^+) \GA(z(\tau),z(\tau^+)) - \chi({\bf x}^-) \GR(z(\tau),z(\tau^-))]
               \, d\tau,
\nonumber \\
  &=& \frac{q}{2}
               [\chi({\bf x}^+) \phitail({\bf x}^+,{\bf x}^+) - \chi({\bf x}^-) \phitail({\bf x}^-,{\bf x}^-)],
\label{scalforcediff}
\end{eqnarray}
where $\phitail({\bf x}_1,{\bf x}_2)$ is the tail part of $\phi$ as
measured at spatial position ${\bf x}_1$ for a point particle source
held stationary at spatial position ${\bf x}_2$ for all time.
Thus, if the particle is initially and finally
stationary at the same spatial position, then the work done by $f_a$
is equal to the work done by $\fg_a$, which is in turn equal to the
energy radiated to infinity by the Klein-Gordon field.

As in the electromagnetic case, eq.~(\ref{scalforcediff}) suggests that
for a particle held stationary at position ${\bf x}$, the renormalized
self-energy stored in the scalar field is given by 
\begin{equation} 
E_{\rm self} ({\bf x}) \equiv \frac{q}{2} \chi({\bf x}) \phi^{\rm tail}
({\bf
x},{\bf x}).  
\label{Eselfscal} 
\end{equation} 
However, in the scalar case, additional energy also is stored in the
mass of the particle. Indeed, from eq.~(\ref{scalarforce}), we obtain
\begin{eqnarray}
\frac{dm}{d \tau} & = & - u^a f_a \nonumber \\ 
& = & - q \frac{d \phi^{\rm in}}{d \tau} - \frac{q^2}{12} R - q^2
\int_{-\infty}^{\tau} \! \frac{d}{d \tau} \GR(\tau,\tau') \,d\tau' \nonumber \\
& = &-  q \frac{d \phi^{\rm in}}{d \tau}  - \frac{q^2}{12} R - q^2
\frac{d}{d \tau} \int_{-\infty}^{\tau} \!\! \GR(\tau,\tau') \,d\tau'
  + q^2 \GR(\tau, \tau)
\label{masschange}
\end{eqnarray}
where in the last line, $\GR(\tau, \tau)$ should be understood to mean
the limit as $\epsilon \rightarrow 0$ of $\GR(\tau,
\tau~-~\epsilon)$. The Hadamard analysis of~\cite{quinnwiseman} shows
that $\GR(\tau,\tau) = R/12$, so we obtain
\begin{equation}
\frac{dm}{d \tau} = - q \frac{d}{d \tau} (\phi^{\rm in} + \phi^{\rm tail})
\end{equation}
Thus, the mass of the particle is now given by
\begin{equation} 
m = m_0 - q(\phi^{\rm in} + \phi^{\rm tail}) 
\end{equation} 
and, hence, the total energy of a stationary particle is
\begin{eqnarray} 
E' &=& - m t^a u_a + E_{\rm self} \nonumber \\ 
&=& [m_0 - q(\phi^{\rm in} + \phi^{\rm tail})] \chi +E_{\rm self}
\nonumber \\ 
&=& E - \frac{q}{2} \chi({\bf x}) \phi^{\rm tail} ({\bf x},{\bf x}) 
\end{eqnarray} 
where $E \equiv (m_0 - q \phi^{\rm in}) \chi$ is the energy that the
particle would have had in the absence of self-force effects. As in
the derivation of eq.~(\ref{Ueff}) in the electromagnetic case, in
general this would lead to a corresponding self-energy correction to
the effective potential. However, in Schwarzschild spacetime, Wiseman
\cite{wiseman} has shown that $\phi^{\rm tail} ({\bf x},{\bf x})$
actually vanishes for all ${\bf x}$, so, to order $q^2$, there is no scalar
self-energy correction to the motion of particle in Schwarzschild
spacetime.

Finally, we note that the analysis of this Appendix can be extended
straightforwardly (in the manner explained in Sec.~II) to treat the
case of angular momentum and to allow for the presence of black holes.

\begin{figure}[ht]
\caption{\label{unphysical} The unphysical spacetime $(\tilde{M},
\tilde{g}_{ab})$ with the hypersurfaces $\tls^+(t^+)$ and
$\tls^-(t^-)$ used in our analysis.}
\end{figure}


\begin{thebibliography}{99}

\bibitem{dirac} P. A. M. Dirac, Proc. Roy. Soc. {\bf A167}, 148 (1938).

\bibitem{dewittbrehme} B. S. DeWitt and R. W. Brehme, Annals of
Physics {\bf 9}, 220 (1960).

\bibitem{hobbs} J. M. Hobbs, Ann. Phys. {\bf 47}, 141 (1968).

\bibitem{minoetal} Y. Mino, M. Sasaki, and T. Tanaka, Phys. Rev. D
{\bf 55}, 3457 (1997).

\bibitem{quinnwald} T. C. Quinn and R. M. Wald, Phys. Rev. D {\bf 56},
3381 (1997).

\bibitem{dewittdewitt} B. S. DeWitt and C. M. DeWitt, Physics (Long
Island City, NY), {\bf 1}, 3 (1964).

\bibitem{quinnwiseman} T. C. Quinn and A. G. Wiseman, in preparation.

\bibitem{jackson} J. D. Jackson, {\it Classical Electrodynamics}, John
Wiley and Sons, (New York, 1975).

\bibitem{galtsov} D. V. Gal'tsov, J. Phys. A {\bf 15}, 3737 (1982).

\bibitem{habisohn} C. X. Habisohn, J. Math. Phys. {\bf 27}, 2759 (1986).

\bibitem{wald1} R. M. Wald, {\it General Relativity}, University of
Chicago Press (Chicago, 1984).

\bibitem{ashtekarhansen} A. Ashtekar and R. O. Hansen,
J. Math. Phys. {\bf 19}, 1542 (1978).

\bibitem{cw} P. T. Chrusciel and R. M. Wald, Commun. Math. Phys. {\bf
163}, 561 (1994).

\bibitem{seifert} H. J. Seifert, Gen. Rel. Grav. {\bf 8}, 815 (1977).

\bibitem{wald2} R. M. Wald, {\it Quantum Field Theory in Curved
Spacetime and Black Hole Thermodynamics}, University of Chicago Press
(Chicago, 1994).

\bibitem{friedman} J. L. Friedman, Commun. Math. Phys. {\bf 63}, 243 (1978).

\bibitem{hormander} L. Hormander, {\it The Analysis of Linear Partial
Differential Operators IV} (Springer, Berlin, 1985).

\bibitem{smithwill} A. G. Smith and C. M. Will, Phys. Rev. D {\bf 22},
1276 (1980).

\bibitem{hubeny} V. Hubeny, gr-qc/9808043.

\bibitem{willwalker} M. Walker and C. M. Will, Phys. Rev. D {\bf 19},
3495 (1979).

\bibitem{garfinkle} D. Garfinkle, private communication.

\bibitem{boulware} D. Boulware, Ann. Phys. (N.Y.) {\bf 124}, 169 (1980).

\bibitem{gx} R. P. Geroch and B. C. Xanthopoulos, J. Math. Phys. {\bf
19}, 714 (1978).

\bibitem{wiseman} A. G. Wiseman, to be published.

\end{thebibliography}
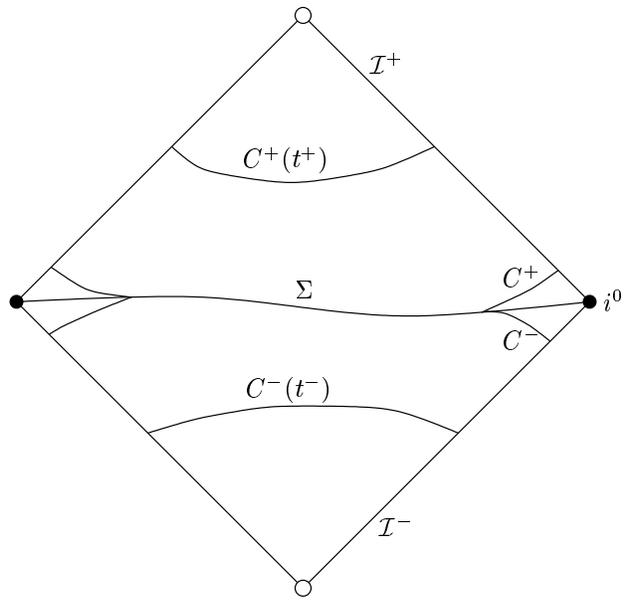
\end{document}